\documentclass[sigconf]{acmart}
\pdfoutput=1
\usepackage{booktabs} % For formal tables
\usepackage{upquote,textcomp}

\usepackage{enumitem}
\usepackage{url}
\usepackage{graphicx}
\usepackage{caption}
\usepackage{color}

\newcommand{\ignore}[1]{}
\newcommand{\qorg}{{\tt Q}$_{o}$\xspace}
\newcommand{\qnew}{{\tt Q}$_{n}$\xspace}
\newcommand{\qbe}{{\tt Q}$_{MV}$\xspace}
\newcommand{\qfe}{{\tt Q}$_{V}$\xspace}
\newcommand{\cfree}{{\tt cfree}\xspace}
\newcommand{\frcc}[1]{\multicolumn{1}{c|}{#1}}
\newcommand{\frccd}[1]{\multicolumn{1}{c||}{#1}}
\newcommand{\frccdo}[1]{\multicolumn{1}{|c||}{#1}}
\newcommand{\gc}{\cellcolor{gray}}
\usepackage{multirow}
\usepackage{array}
\newcolumntype{L}{>{\centering\arraybackslash}m{1.5cm}}
\usepackage{xspace}
\usepackage{fancyvrb}
\usepackage{algorithm}
\usepackage{algorithmic}
\usepackage{float}
\floatstyle{ruled}
\newfloat{myalgo}{thp}{lop}
\floatname{myalgo}{Algorithm}
\usepackage{alltt}
\usepackage{fancyvrb}

\setcopyright{none}
\renewcommand\footnotetextcopyrightpermission[1]{} % removes footnote with conference information in first column
\makeatletter
\renewcommand\@formatdoi[1]{\ignorespaces}
\makeatother

\begin{document}
\title{In-Browser Split-Execution Support for\\ Interactive Analytics in the Cloud}

\author{Kareem El Gebaly and Jimmy Lin}
\affiliation{
  \vspace{0.1cm}
  David R. Cheriton School of Computer Science\\
  University of Waterloo
}
\email{{kareem.elgebaly, jimmylin}@uwaterloo.ca}

\fancyhead{}
\settopmatter{printacmref=false, printfolios=true}

\begin{abstract}
The canonical analytics architecture today consists of a browser
connected to a backend in the cloud. In all deployments that we are
aware of, the browser is simply a dumb rendering endpoint. As an
alternative, this paper explores split-execution architectures that
push analytics capabilities into the browser. We show that, by taking
advantage of typed arrays and asm.js, it is possible to build an
analytical RDBMS in JavaScript that runs in a browser,
achieving performance rivaling native databases. To support
interactive data exploration, our Afterburner prototype automatically
generates local materialized views from a backend database that are
then shipped to the browser to facilitate subsequent interactions
seamlessly and efficiently. We compare this architecture to several
alternative deployments, experimentally demonstrating performance
parity, while at the same time providing additional advantages in
terms of administrative and operational simplicity.
\end{abstract}

\maketitle

\section{Introduction}

For today's data scientists, the browser has become the shell,
especially for interactive analytics on large datasets. What used to
be accomplished via command-line REPL-based tools is increasingly
moving into browser-based notebooks such as Jupyter. Such tools have
gained popularity due to their first-class support for interactive
data exploration. Notebooks also tie analytics tools into existing
software ecosystems (e.g., Python, R), giving data scientists access
to a broad range of capabilities. At the same time, backend analytics
capabilities are increasingly centralized in the cloud, exemplified by
various database-as-a-service offerings (Amazon RDS, Azure SQL, etc.),
fully-managed data analytics stacks (Google BigQuery), as well as more
general dataflow frameworks (Google Cloud Dataflow, Amazon Spark EMR).

Given these two trends, the canonical client--server architecture today
consists of a browser connected to a cloud backend. In all
implementations that we are aware of, the browser is simply a dumb
rendering endpoint:\ all query execution is handled by backend
servers. However, modern browsers are capable of much more:\ they
embed powerful JavaScript engines capable of running real-time
collaborative tools, rendering impressive 3D scenes, and even running
first-person shooters.

We asked:\ Is it possible to exploit modern JavaScript engines to
rethink the design of current architectures? In particular, would it
be possible to offload data management and analytics capabilities from
the cloud into the browser, and if so, what advantages might such an
architecture offer? We explore the possibilities in this paper.

\smallskip \noindent {\bf Contributions.} 
We view our work as having the following two main contributions:

\begin{itemize}[leftmargin=*]

\item Building on a previous demonstration~\cite{ElGebaly_Lin_SIGMOD2017}, we describe and evaluate
  Afterburner, our prototype analytical RDBMS implemented in JavaScript
  that runs completely in the browser. We detail how our system takes
  advantage of in-memory columnar storage using typed arrays and query
  compilation into asm.js. Microbenchmarks, as well as end-to-end
  evaluations, show that our techniques approach the performance of
  an existing columnar database (MonetDB) and modern query
  compilation techniques (LegoBase) running natively. While we are
  not the first to implement an SQL engine in JavaScript, our
  prototype offers far superior scalability and performance on
  analytical queries compared to alternatives such as Google's
  Lovefield and Sql.js.

\item To support a large class of interactive SQL analytics, we
  propose a novel technique whereby the data scientist provides a hint
  regarding the focus of data exploration, and our system
  automatically splits query execution across the backend and the
  browser. The backend generates a local materialized view that is
  shipped over to the browser for subsequent manipulation. Note
  that we do not claim significant innovations in
  materialized view techniques. Instead, our contribution lies in a novel
  approach to integrating analytics backends with in-browser
  processing in a deployment that has many attractive features. We
  compare a number of deployment architectures and empirically
  demonstrate the advantages of our design.

\end{itemize}

\section{Background}
\label{section:background}

The work of data scientists often involves tight interaction cycles
with the data, particularly for exploratory tasks. This paper focuses
on and optimizes for the common scenario where a data scientist
rapidly issues a sequence of SQL queries that differ only in the
predicates in the \texttt{WHERE} clause. As a running example,
consider Q6 from the TPC-H benchmark for decision support
in data warehousing:

\smallskip
\begin{small}
\begin{verbatim}
  SELECT SUM(l_extendedprice * l_discount) AS revenue
  FROM   lineitem
  WHERE  l_shipdate >= DATE '1994-01-01'
     AND l_shipdate < DATE '1995-01-01'
     AND l_discount BETWEEN 0.05 and 0.07
     AND l_quantity < 24;
\end{verbatim}
\end{small}

\smallskip

\noindent We quote from the benchmark definition:
``This query quantifies the amount of revenue increase that would have
resulted from eliminating certain companywide discounts in a given
percentage range in a given year. Asking this type of `what if'
query can be used to look for ways to increase revenues.''
From this, we can see that as part of data exploration,
a data scientist might be interested in the results of the same
``query template'', but with different predicates on the
\texttt{WHERE} clause:\ different dates, discounts, etc.

Further evidence in support of the prevalence of such queries can
be found in the more recent TPC-DS benchmark, which explicitly talks
about query templates to test the interactive and iterative nature
of OLAP queries (see~\cite{Poess:2007:WYR:1325851.1325979}, Section~4). 
An example of an iterative query, Q24, captures drill down and
the iterations vary only in the \texttt{WHERE} predicates. This matches
exactly the scenario we envision. We note that previous work such as
BlinkDB~\cite{Agarwal:2012:BDI:2367502.2367533} also makes the
assumption that query templates are relatively stable, which is the
key to effective sampling. Similarly, Verdict~\cite{verdict} assumes
a stream of queries with correlated results, which necessarily must share
commonalities, as the basis of learning an underlying model for approximate
query answering; the work of Galakatos et al.~\cite{alexg} exploits similar intuitions as well.
Agarwal et al.~\cite{Agarwal:2012:BDI:2367502.2367533} further cite evidence
from production workloads that support such query behaviors. In
fact, dashboards and report generators are essentially pre-specified
query templates:\ another way to look at our work is that it
supports the creation of interactive dashboards on demand given ad
hoc queries. Although our techniques certainly do not cover all
the activities of a data scientist, we believe that we are tackling a
common and realistic use case.

\begin{figure}[t]
  \includegraphics[width=.45\textwidth, keepaspectratio]{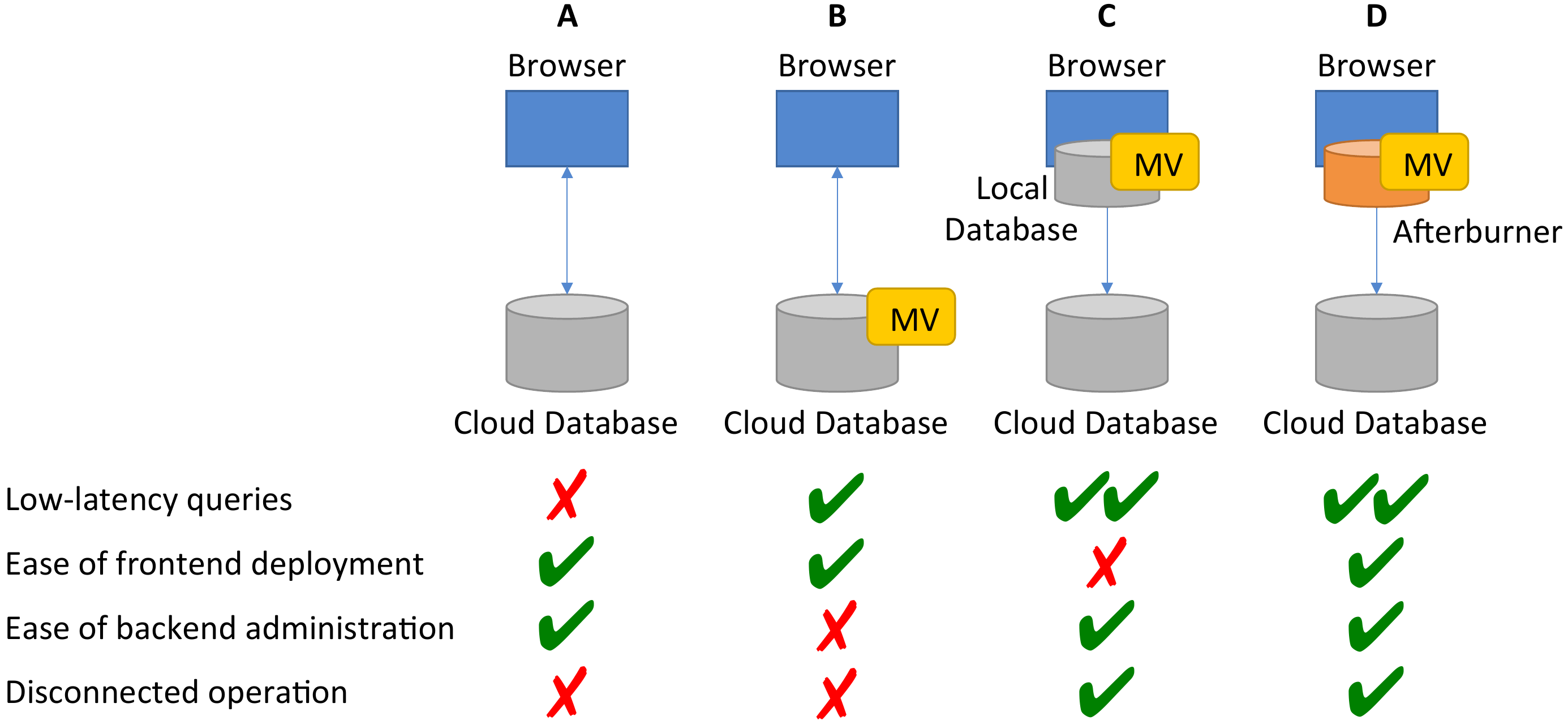}
  \caption{Tradeoffs between various deployment scenarios to support
    interactive analytics in the cloud.}
  \label{fig:deployment}
\end{figure}

To support such data exploration, we desire
three important properties:\ that such queries be fast (i.e., low
latency), simplicity of frontend deployment (minimizing the effort of
data scientists in managing their own client setup), and simplicity of
backend administration (minimizing the effort of data warehouse
administrators to support data scientists). With this in mind, let us
consider a number of deployment scenarios, summarized in
Figure~\ref{fig:deployment}:

\smallskip \noindent {\bf Deployment A.} This is the basic browser
client connected to a cloud analytics backend (for concreteness, an
analytical RDBMS). With this setup, the data scientist re-issues the
same query template with different bindings, leading to sub-optimal
query latencies because each SQL query is treated
independently.

\smallskip \noindent {\bf Deployment B.}  An obvious improvement to
Deployment A is to accelerate analytics with a materialized view (MV),
either explicitly requested by a data scientist or automatically
inferred via query rewriting.  Such an optimization does not alter how
the client is deployed (so we retain simplicity in frontend
deployment) but introduces new administrative burdens on the
backend. In terms of creating materialized views:\ Are all data
scientists able to create materialized views? Are there quotas to
ensure equitable use of resources? Are these views ``transient'' or
can they persist over long periods? If the former, what are the
mechanisms by which these views are garbage collected?  Are these
views maintained as new data arrive?  If so, what is the impact on
system performance? If no, how are these views invalidated?  Although
we have improved query performance, we have sacrificed simplicity
of backend administration in terms of a number of policy and
technical issues that must be addressed.

\smallskip \noindent {\bf Deployment C.}  Another improvement to
Deployment A is to take the materialized view and move it over to the
client, stored in a local RDBMS. 
This approach eliminates the backend administration issues described
in Deployment B because data scientists can manage their own machines.
This approach, however, complicates frontend deployment. For example,
how does the entire team keep up to date with the
latest version of the local RDBMS? What if there are urgent security
patches that need to be installed?  What about different operating
systems and other idiosyncratic configurations
across many heterogeneous machines in
an enterprise setting? In fact, the headache of managing these issues
is what drove organizations to the cloud to begin with, so Deployment
C seems like a step backward.

\smallskip \noindent {\bf Deployment D.}  The central value
proposition of this work is that we can have our cake and eat it
too. Building on Deployment C, what if the ``local'' database were
written in JavaScript and runs in the browser? In that case,
``deployment'' is as simple as loading a webpage. This is Afterburner.
The obvious question is, of course, how much performance are we giving
up with a JavaScript analytical RDBMS? The answer, as we
experimentally show, is nothing:\ end-to-end query latency is
comparable to Deployment C.

We note that Deployment D has two additional advantages:\ First, it
supports disconnected operation when access to the cloud is
unavailable. 
Deployment C also offers this flexibility, but not A or
B. Second, Deployment D supports multi-device deployment, e.g., on
tablets and even mobile phones---since it's just JavaScript. 
This feature is not supported by Deployment C, since a local
database may not be available on all devices.

\section{Afterburner Design}
\label{section:design}

\noindent We begin by detailing the design of Afterburner, which takes
advantage of two JavaScript features:\ typed arrays for
memory-efficient storage and asm.js for fast compiled queries. In this
section, we focus on Afterburner as a standalone, in-browser analytical
RDBMS.

\subsection{Columnar Storage with Typed Arrays} 

Array objects in JavaScript can store elements of any type and are not
arrays in a traditional sense (compared to say, C) since consecutive
elements may not be contiguous; furthermore, the array itself can
dynamically grow and shrink. This flexibility limits the optimizations
that the JavaScript engine can perform both during compilation and at
runtime. In the evolution of JavaScript, it became clear that the
language needed more efficient methods to manipulate binary
data: typed arrays are the answer.

Typed arrays in JavaScript are comprised of buffers, which simply
represent untyped binary data, and views, which impose a read context
on the buffer. As an example, the following creates a 64-megabyte buffer:

\smallskip
\begin{small}
\begin{Verbatim}[commandchars=\\\{\}]
  var heap = new ArrayBuffer(64*1024*1024);
\end{Verbatim}
\end{small}
\smallskip

\noindent Before we can manipulate the data, we need to create a view
from it. With the following:

\smallskip
\begin{small}
\begin{Verbatim}[commandchars=\\\{\}]
  var hI32 = new Int32Array(heap);
\end{Verbatim}
\end{small}
\smallskip

\noindent we can now manipulate {\tt hI32} as an array of 32-bit
integers (e.g., iterate over it with a {\tt for} loop).

\begin{figure}[t]
  \includegraphics[width=.48\textwidth, keepaspectratio]{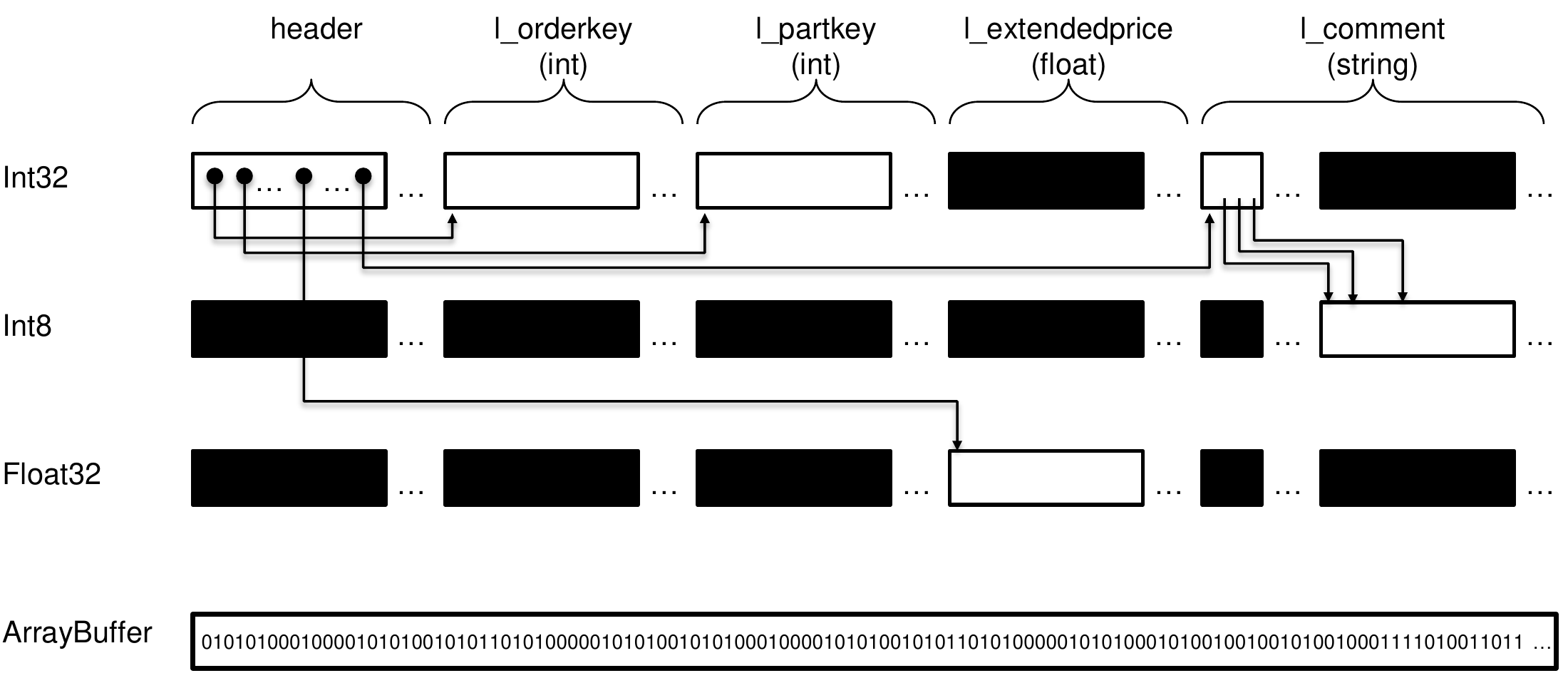}
  \caption{Illustration of the physical in-memory layout of
    the \texttt{lineitem} table from TPC-H. Different ``views'' (top
    three rows) provide access into the underlying JavaScript
    ArrayBuffer, which holds the actual data (bottom row).
    Blackened boxes represent invalid data for that
    particular view.}
  \label{fig:memory}
\end{figure}

Typed arrays allow the developer to create multiple views over the
same buffer. Afterburner takes advantage of this feature to pack
relational data into a columnar layout.
In our implementation, each column is laid out end-to-end in the underlying
buffer, which can be traversed with a view of the corresponding
type. The table itself is a group of pointers to the offsets of the
beginning of the data in each column.  Figure~\ref{fig:memory} shows
the physical memory layout storing the \texttt{lineitem} table from the
TPC-H benchmark, which we use as a running example.  A \texttt{lineitem}
pointer serves as the entry point to a group of 32-bit integer
pointers, which represent the offsets of the data in each column ({\tt
  l\_orderkeys}, {\tt l\_partkeys}, etc.).  Currently, Afterburner
supports integers, floats, dates, and strings.  For the
first three types, values are stored as literals (essentially, as an
array). For a column of strings, we store null-terminated strings
prefixed with a header of pointers to the beginning of each string,
essentially a {\tt (char **)} in C. In Figure~\ref{fig:memory},
the bottom row represents the raw buffer, while the top three rows represent
the various views. Blackened boxes represent invalid data for that particular view,
and can be interpreted as masks for a particular column. Although data is not compressed
in memory, there is no principled reason why various columnar compression
techniques~\cite{Abadi_etal_SIGMOD2006} cannot be applied in our implementation.
Intermediate data for query execution in Afterburner are also stored
using typed arrays.

In this paper, we adopt the following naming convention for
JavaScript variables:\ {\tt heap} refers to an instance of a binary
ArrayBuffer. Names of views over the typed array begin with an {\tt
  h}, followed by the first letter of the data type, then the size of
the data type in bits. For example, {\tt hI8}, {\tt hI32}, and {\tt
  hF32} represent Int8, Int32, and Float32 views over a typed
array.

\subsection{Query Compilation into Asm.js} 
\label{s:qcasm}

In conjunction with typed arrays, Afterburner takes advantage of asm.js, a
strictly-typed subset of JavaScript that is designed to be easily
optimizable by an execution engine. Consider the following fragment of
JavaScript for counting the number of records that matches a
particular predicate on the {\tt l\_extendedprice} column:\footnote{We
  refer to JavaScript without asm.js optimizations as {\it vanilla}
  JavaScript.}

\smallskip
\begin{small}
\begin{verbatim}
  function count(val){                  
    var cnt = 0;
    for(var id = 0; id < l_extendedprice.length; id++)
      if (l_extendedprice[id] < val) cnt++;
    return cnt;
  }                                      
\end{verbatim}
\end{small}
\smallskip

\noindent The equivalent function in asm.js is as follows:

\smallskip
\begin{small}
\begin{Verbatim}[commandchars=\\\@\#]
  function count_asm(val, l_extendedprice, length){
    "use asm";
    val =+ (val);
    length = length|0;
    length = length << 2;
    id = 0;
    while((id|0) < (length|0)){
      if(+(hF32[((l_extendedprice + id)|0) >> 2]) < +(val))
        cnt = (cnt + 1)|0;
      id = (id + 4)|0;
    }
    return cnt|0;
  }
\end{Verbatim}
\end{small}
\smallskip

\noindent The above function takes as parameters: 
{\tt l\_extendedprice}, which is the starting byte offset
of the {\tt l\_extendedprice} column, and {\tt length}, which is
the number of records in that table.
The {\tt hF32} variable is a 32-bit float view and
thus the byte offsets can be computed by multiplying the index variable
{\tt id} by four using the shift operator ({\tt <{<}}).
Asm.js uses type hints, such as
{\tt x|0} and {\tt +(x)}, which are applied to variables or
arithmetic expressions. The type hint ({\tt x|0}) specifies a 32-bit
integer and {\tt +(x)} specifies a 32-bit floating point value.  With
these hints, asm.js essentially introduces a static type system while
retaining backwards compatibility, since in
``vanilla'' JavaScript these hints just become no-ops.  

Any JavaScript function can request validation of a block of code as
valid asm.js via a special prologue directive, {\tt use asm}, which
happens when the source code is loaded.  Validated asm.js code
(typically referred to as an asm.js module) is amenable to
ahead-of-time (AOT) compilation, in contrast to just-in-time (JIT)
compilation in vanilla JavaScript. Executable code generated by AOT
compilers can be quite efficient, through the removal of runtime type
checks (since everything is statically typed), operations on unboxed
(i.e., primitive) types, and removal of garbage collection.

An asm.js module can take three optional parameters, which
provide hooks for integration with external JavaScript
code:\ a standard library object, providing access to a limited subset
of the JavaScript standard libraries; a foreign function interface
(FFI), providing access to custom external JavaScript functions; and a
heap buffer, providing a single ArrayBuffer to act as the asm.js
heap.\footnote{\url http://asmjs.org/spec/latest/} Thus, a typical
asm.js module declaration is as follows:

\smallskip
\begin{small}
\begin{verbatim}
  function AsmModule(stdlib, ffi, heap){
    "use asm";
    // module body
  }
\end{verbatim}
\end{small}
\smallskip

\noindent At a high-level, Afterburner translates SQL into the string
representation of an asm.js module (i.e., the physical query plan,
through code templates described below), calls {\tt eval} on the code,
which triggers AOT compilation and links the module to the calling
JavaScript code, and finally executes the module (i.e., executes the
query plan). The typed array storing all the tables 
is passed into the module as a parameter, and the query
results are returned by the module.

Instead of string-based SQL queries, Afterburner executes queries
written using an API that is heavily driven by method chaining, often
referred to as a {\it fluent} API.
There is a straightforward mapping from the method calls to
clauses in an SQL query, so we can view the fluent API as
little more than syntactic sugar. However, this query API
is quite similar to
DataFrames~\cite{Armbrust_etal_SIGMOD2015}, an interface for data
manipulation that many data scientists are familiar with today.
In fact, Lovefield, one of the SQL-in-JavaScript systems we
discuss in Appendix~\ref{sec:existing}, adopts the same style
for specifying SQL queries.

Starting from an SQL query expressed in our fluent API,
Afterburner generates the string representation of the asm.js code
that corresponds to the query. In the current implementation, this is
performed based on a small number of fixed
code templates in which various sub-expressions (e.g., the filter
predicate, join key, group by clause, etc.)\ are plugged. At present,
Afterburner has a fixed (hard-coded) physical plan for each class of
queries (i.e., it does not perform query optimization). Our current
implementation supports a wide variety of SQL analytical operations,
covering all the queries in the TPC-H benchmark:

\begin{itemize}[leftmargin=*]

\item {\bf Simple Filters.} We have a basic code template that
generates query plans for simple filter--project or filter--aggregate
queries. The template generates a loop that increments a record
iterator, which is used in combination with the starting offset of a
column to access a particular attribute. Inside the loop, the template
can either generate code to materialize a projection or to compute
simple aggregates such as {\tt COUNT}, {\tt AVG}, or {\tt SUM}.

\item {\bf Joins.} The code template for supporting
filter--project or filter--aggregate queries over an inner join
implements a standard hash join. In the build phase, the code loops
over one relation to build the hash table. In the probe phase, the
generated code loops over the second relation to probe the hash table
for matching records, and then either materializes a projection or
computes an aggregate. The template currently only supports two-way
joins.

\item {\bf Group Bys.} The code template loops over one
relation to build a hash table over the grouping keys. Another
loop iterates over the hash table in order to process the groups.

\item {\bf Subqueries.} Afterburner handles subqueries
by materializing their output, which is used as input tables for other
operators. In addition, Afterburner supports {\tt IS IN} subquery
clauses by generating code that is similar to the join code templates.

\end{itemize}

\subsection{In-Browser Query Processing}

In the previous sections, we have outlined two main features that
characterize Afterburner's SQL engine, which are columnar
storage and code generation. Other than the challenges associated
with the JavaScript runtime, the web browser imposes additional challenges
such as a limited memory footprint. In this section, we discuss
design decisions behind Afterburner's code generation templates that 
allow running query operators efficiently inside the browser. 

\begin{figure}[t]
\begin{center}
\begin{small}
\begin{Verbatim}[commandchars=\\\@\#]
 1:while(1){l_rid=l_rid+1|0;
 2:  if ((l_rid|0) >= 6000000) break;
 3:  hk=((((hI32[(oOffset+ (l_rid<<2)) >>2]|0)));
 4:  hk= hk &(hashBitFilter|0))|0;
 5:  bucket=-1; curr=0;
 6:  if (hI8[(h1db + (hk>>3))|0] & (1<<(hk&7)))
 7:    bucket=hI32[((h1bb+(hk<<2))|0)>>2]|0;
 8:  while(((bucket|0)>0)){
 9:    if ((curr|0)>=(hI32[bucket>>2]|0)){
10:      bucket=hI32[(bucket+(((h1Size+1)|0)<<2)|0)>>2]|0;
11:      if (bucket){
12:        curr=1;
13:        o_rid=hI32[((bucket+(curr<<2))|0)>>2]|0;
14:      } else
15:        break;
16:    } else {
17:        curr=curr+1|0;
18:        o_rid=hI32[((bucket+(curr<<2))|0)>>2]|0;
19:    }
20:    if (!((+((hI32[((oOffset+(l_rid<<2))|0)>>2]|0)|0))
21:        ==(+((hI32[((iOffset+(o_rid<<2))|0)>>2]|0)|0))));
22:      continue;
23:    sum1=sum1+(+(hF32[((8196588 +(l_rid<<2))|0)>>2]));
24:  }
25:}
\end{Verbatim}
\end{small}
\end{center}
\vspace{-0.1cm}
\caption{Sample asm.js compiled query fragment.}
\label{f:joincode}
\end{figure}

\smallskip \noindent {\bf Copy-free joins and group by code
  templates.}  Afterburner avoids materialization of intermediate results
as much as possible in order to run inside a web browser. For example,
it only stores record identifiers in hash tables (instead of copying
the values themselves) in order to minimize the memory footprint of
the hash-based join and group by operators. Afterburner will translate
record identifiers into array indices to access the column values in a
lazy fashion. An alternative design that stores record values directly
in the hash table can provide faster performance during the probe
phase by avoiding the extra overhead required to retrieve the values
themselves (e.g., array index translation and random memory access).
However, Afterburner's lazy approach requires less memory and avoids
unnecessary copying.

As an example, consider the following query, expressed in
fluent SQL, which computes the sum of {\tt l\_extendedprice} column
from the \texttt{lineitem} table, filtered by a certain order date range.
In order to apply this filter, a join with the \texttt{orders} table is
required:

\smallskip
\begin{small}
\begin{verbatim}
  abdb.select()
    .from('lineitem')
    .join('orders')
    .on('l_orderkey','o_orderkey')
    .field(sum('l_extendedprice'))
    .where(lt('o_orderdate', date('1995-03-15')));
\end{verbatim}
\end{small}
\smallskip

\noindent Figure~\ref{f:joincode} shows the code generated by the
template for the join operator in Afterburner. We
  assume a hash table on the {\tt o\_orderkey} records that satisfy
  the predicate on the {\tt o\_orderdate} column has already been built at an
  earlier stage.
The hash table and the column storage use JavaScript typed array views
over a single buffer ({\tt heap}).
Since the query does not have any filter predicates over the \texttt{lineitem}
table, lines 1 and 2 control a loop over the {\tt l\_orderkey} 
column from zero up to the number of records in the table. 
The variable {\tt l\_rid} is used to maintain this record identifier,
which is then translated into an array index with the relevant
data type in order to access the values of the column, as described
in Section~\ref{s:qcasm}. Lines 3 and 4 compute hash keys of the
values in the {\tt l\_orderkey} column, which are then used to
look up record identifiers of the matching records in {\tt o\_orderkey}.

Next, lines 6--19 are responsible for probing the hash table over the
{\tt o\_orderkey} column. 
A matching hash key does not guarantee that the values
of {\tt o\_orderkey} and {\tt l\_orderkey} match, since 
hash values might collide in false positives. Thus, lines 
20--22 use the record identifier {\tt l\_rid} and {\tt o\_rid},
which was retrieved by probing the hash table, to compute array
indices in order to compare the actual values. Finally, line 23 uses
the record identifier for the \texttt{lineitem} table to compute the
memory address of the associated {\tt l\_extendedprice} value in
order to compute the sum.

\smallskip \noindent {\bf Hash table implementation.}
Our hash table uses chaining for collision resolution, where
the chains are allocated as unrolled linked lists in another memory segment.
Since we do not have access to operating system calls such as {\tt malloc}
and {\tt free}, Afterburner must handle all aspects of memory management itself.
In our example compiled query, the variable {\tt h1bb} shown in line 7 holds the starting array
index of the hash table segment. At the beginning of a new query,
the generated code must reset the hash table, i.e., 
remove any previously-inserted keys. In our experiments, we found 
that resetting hash table segments takes substantial time; for example,
setting the value of 2.6 million keys (2.6 million $\times$ 4 byte integers = 
10MB of memory) takes on average 9ms on our client machine, which is an 
unacceptable overhead per operator usage, since an operator might be used multiple
times per query. 

To minimize the overhead associated with resetting the hash table after each use
(i.e., zeroing out the memory), we maintain an array of bits that
tracks the state of each hash table key. 
Thus, a hash table for 2.6 million keys requires
a bit array of only 300KB and takes less than 1ms to reset.
To illustrate this, consider the code shown in Figure~\ref{f:joincode}:\
lines 6--8 check whether the bit associated with the
key is set to one or zero before checking whether the key exists in the 
hash table. Only if the associated bit is set to one does the
system consider the key to be valid. This allows hash-based operators to
reuse memory segments with little overhead.

Another benefit of the bit array is when using the hash table for group
by operators on sparse columns 
(columns with a small number of unique values relative to the
cardinality of the relation, for example, region identifiers in the TPC-H schema).
After the insertion phase is complete,
the generated code first scans the keys stored in the hash table in order to 
check for unique values (the groups). Scanning the bit array can uncover
the inserted keys in a fraction of the time required to scan the entire
array segment that stores the actual keys. For dense columns (columns
with a large number of unique values relative to the size of the relation),
checking the bit array does not add much overhead.

\section{Split Execution} 
\label{section:split}

As discussed, our work tackles the common scenario in
interactive SQL data exploration where a data scientist executes a
sequence of queries that differ only in the predicates in the
\texttt{WHERE} clause. Our running example Q6 from the TPC-H benchmark, which
examines ``what-if'' revenue missed because of discounts, exemplifies
this scenario. The data scientist might be interested in exploring
revenue missed under different date ranges (i.e., different predicates
on the {\tt l\_shipdate} column). 
Instead of issuing a different SQL query each time,
she can use the following materialized view to answer this query
for all date ranges:

\smallskip
\begin{small}
\begin{Verbatim}[commandchars=\\\{\}]
  CREATE MATERIALIZED VIEW Q6MVsql AS (
  SELECT SUM(l_extendedprice * l_discount) AS f1, 
         l_shipdate
  FROM   lineitem
  WHERE  l_discount BETWEEN 0.05 and 0.07
    AND  l_quantity < 24
  GROUP BY l_shipdate );
\end{Verbatim}
\end{small}
\smallskip

\noindent The materialized view has two columns, one with precomputed sums
({\tt f1}) grouped by {\tt l\_shipdate} and the other with the associated
{\tt l\_shipdate}. Q6 can be computed using
this view by applying the filter predicate to the \texttt{l\_shipdate}
column and computing a sum over the precomputed sums ({\tt f1}), expressed as follows:

\smallskip
\begin{small}
\begin{Verbatim}[commandchars=\\\{\}]
  SELECT SUM(f1) AS revenue
  FROM   Q6MVsql
  WHERE  l_shipdate >= DATE '1994-01-01'
     AND l_shipdate < DATE '1995-01-01';
\end{Verbatim}
\end{small}
\smallskip

\noindent Afterburner is able to automatically and transparently rewrite
an SQL query into two separate queries based on a hint provided by
the data scientist. We introduce a \texttt{FREE} clause that 
allows the data scientist to specify the column over which she
wishes to issue follow-up SQL queries with different predicates
(\texttt{l\_shipdate} in this case). Afterburner
then generates the two queries above:

\begin{list}{\labelitemi}{\leftmargin=1em}

\item The {\bf materialized view query} or \qbe, which builds the
  appropriate materialized view.

\item The {\bf view query} or \qfe, which is the new query rewritten
  against the materialized view.

\end{list}

\noindent The data scientist can express Q6 using fluent SQL,
  using the \texttt{free} operator on the {\tt l\_shipdate} column as
  follows:

\smallskip
\begin{small}
\begin{Verbatim}[commandchars=\\\{\}]
  Q6MVjs = abdb.select()
    .from('lineitem')
    .field(as(sum(mul('l_extendedprice', 'l_discount')),
             'revenue'))
    .where(between('l_discount', 0.05, 0.07))
    .where(lt('l_quantity', 24))
    .free('l_shipdate');
\end{Verbatim}
\end{small}
\smallskip

\noindent Calling \texttt{free} in the last line runs the materialized
view query at the backend and copies it to the frontend.
After this, the data scientist can interactively explore the data by adding more filters on the
{\tt l\_shipdate} column (let's call it {\tt Q6js}), which 
can be executed in the frontend by calling the {\tt exec}
function.

\smallskip
\begin{small}
\begin{Verbatim}[commandchars=\\\{\}]
  Q6js.exec(Q6MVjs);
\end{Verbatim}
\end{small}
\smallskip

\noindent The parameter {\tt Q6MVjs} is passed to the {\tt exec} function as
the materialized view to use.

This scenario illustrates how Afterburner
creates a materialized view transparently using our API, i.e.,
the data scientist did not have to come up with the SQL query definition 
for the materialized view. In addition, she did not have to
rewrite her query against the materialized view. This is 
desirable because she can work with multiple materialized
views at the same time, working from the same query template with
minimal modifications.
We emphasize that these materialized views are {\it local}
with respect to the data scientist and not updated as the
original data sources change. In our usage scenario, the views
are intended to be transient and lightweight; Section~\ref{view:discussion} discusses this in more detail.

\subsection{Materialized View Query}

\begin{figure}[t]
  \includegraphics[width=.4\textwidth, keepaspectratio]{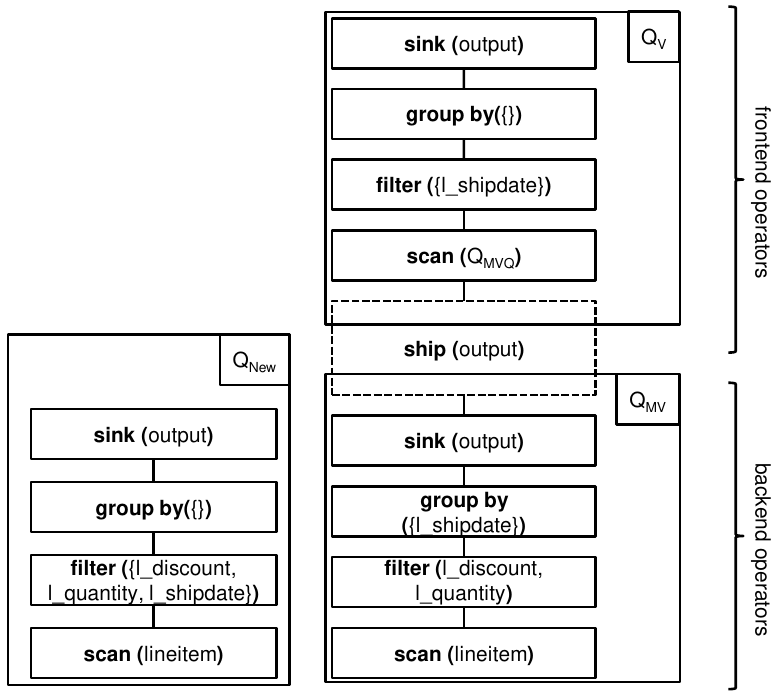}
  \caption{The original (left) and split-execution (right) query plans for Q6.}
  \label{fig:spex}
\end{figure}

Figure~\ref{fig:spex} illustrates the intuition of how Afterburner splits
the execution of a query. On the left, we show a simple plan for Q6,
starting with a {\tt scan} operator that produces records based on the
projected columns from the \texttt{lineitem} table. This is followed
by a {\tt filter} operator, which applies the predicates to the records,
then a {\tt group by} operator that applies the requested
aggregation function. Finally, the {\tt sink}
operator produces the output. Although we can materialize the output
of any operator, save it, ship it to the browser, and then apply
the rest of the operators, this may not be sufficient to ``free up''
the column.

To free the {\tt l\_shipdate} column, we transform the query in a way that
delays the filter (shown on the right side
of Figure~\ref{fig:spex}). This transformation requires passing the  
{\tt l\_shipdate} column to the {\tt group by} operator. Next, 
we materialize the output and ship it to the browser and then apply
the rest of the plan. When the user changes the predicates
on the {\tt l\_shipdate} column, only the local part of the plan
must be changed---and can be executed in the browser using Afterburner.

Now that we have illustrated the intuition behind the materialized
view generation, we discuss how Afterburner generates the SQL
definition of the materialized view. Targeting SQL to create 
materialized views (instead of a physical plan, for example)
comes with a few benefits, which includes widening the applicability 
of our techniques to any backend that supports SQL (e.g., Spark). 
In addition, using SQL allows the backend to adopt the best plan
according to its own query optimizer to answer the query.  

\begin{myalgo}[t]
\begin{alltt}
\textnormal{
\textbf{Input: }\texttt{Query} \qorg, \texttt{Column} \cfree
\texttt{ 1:}{  }\qbe \(\leftarrow\) \textbf{new} \texttt{Query}()
\texttt{ 2:}{  }\qbe.\texttt{SELECT} \(\leftarrow\) \qorg.\texttt{SELECT} \(\cup\) \texttt{\{cfree\}}
\texttt{ 3:}{  }\qbe.\texttt{FROM} \(\leftarrow\) \qorg.\texttt{FROM}
\texttt{ 4:}{  }\qbe.\texttt{GROUP} \(\leftarrow\) \qorg.\texttt{GROUP}
\texttt{ 5:}{  }\qbe.\texttt{WHERE} \(\leftarrow\) \qorg.\texttt{WHERE} \(\setminus\) \texttt{\{cfree\}}
\texttt{ 6:}{  }\textbf{for} c \(\in\) \qbe.\texttt{SELECT} \(\land\) \texttt{isAggregate}(c):
\texttt{ 7:}{      }c \(\leftarrow\) \texttt{rewriteAgg}(c)
\texttt{ 8:}{      }hasAggregate \(\leftarrow\) true
\texttt{ 9:}{  }\textbf{if } hasAggregate:
\texttt{10:}{      }\qbe.\texttt{GROUP} \(\leftarrow\) \qbe.\texttt{GROUP} \(\cup\) \texttt{\{cfree\}}
\texttt{11:}{  }\textbf{return} \qbe}
\end{alltt}
\caption{Generate MVQ}
\label{algfree}
\end{myalgo}

\begin{table}[t]
\begin{center}
\begin{small}
\resizebox{0.47\textwidth}{!}{
\begin{tabular}{|l|l|}
\hline
{\bf Symbol}     & {\bf Explanation}\\ \hline \hline
\qorg            & Original query with a \texttt{FREE} clause. \\ \hline
\cfree           & A column set to be free in the \texttt{FREE} clause of \qorg. \\ \hline
\qbe             & Materialized view query which is the SQL definition of the materialized view. \\ \hline
\qnew            & A new query submitted with a corresponding \qbe to use. \\ \hline
\qfe             & A new query generated by Afterburner to run in the frontend. \\ \hline
\multirow{3}{*}{{{\tt Q}}.SELECT}  &The list of terms in the {\tt SELECT} clause of query {{\tt Q}}. A term in the {\tt SELECT} \\ 
                 & clause, can be a constant value, an attribute name, or an aggregation\\
		 & function over an attribute name.  \\ \hline
\multirow{2}{*}{{{\tt Q}}.FROM} & The list of terms in the {\tt FROM} clause of query {{\tt Q}}. A term in the {\tt FROM}      \\ 
                 &  clause can be a table name or a subquery. \\\hline
{{\tt Q}}.WHERE  & The list of predicates in the {\tt WHERE} clause of query {{\tt Q}}. \\ \hline
{{\tt Q}}.GROUP  & The list of columns in the {\tt GROUP BY} clause of query {{\tt Q}}. \\ \hline
\end{tabular}
}
\end{small}
\end{center}
\vspace{0.2cm}
\caption{Explanation of symbols used in this section.}
\label{t:symbols}
\vspace{-0.2cm}
\end{table}

Materialized view query generation is shown in Algorithm~\ref{algfree}, which computes the
materialized view definition (\qbe) based on an original query 
(\qorg) and a valid column to free (\cfree). 
Symbols used here and in the following descriptions are summarized in Table~\ref{t:symbols}.
Line 1 initializes an
empty query template (\qbe{}). Line 2 adds each term in the 
{\tt SELECT} list of \qorg{} to \qbe{}. For example, in Q6 the 
select clause of the original query contains the aggregated value
{\tt SUM(l\_extendedprice * l\_discount)} which is added to the 
materialized view query. In addition, line 2 adds the column
in the {\tt FREE} clause to the {\tt SELECT} list of \qbe{}, which is
{\tt l\_shipdate}.
This step is necessary to be able to apply predicates on
\cfree\ at the frontend. Lines 3--4 add the terms in the
{\tt FROM} and {\tt GROUP BY} lists of \qorg{} to \qbe{}. The terms
in the \texttt{FROM} clause can be any valid term, i.e., a table name or
a subquery.

In principle, the materialized view \qbe{} may inherit filter predicates
on the \cfree\ column from the original query \qorg{}, which limits
subsequent queries.
For example, the original query can specify a certain
date range, which will limit subsequent queries on \texttt{l\_shipdate} to be within that date range.
However, in our implementation, we employ a simpler approach and do not consider
filters on \cfree, which simplifies the {\tt FREE} clause. Thus, line 5
removes any instance of \cfree\ from the {\tt WHERE} list.

Lines 6--8 check for aggregates in the {\tt SELECT} list of 
\qbe{} and rewrite the aggregation functions based on predefined rules.
For example, we rewrite 
{\tt AVG} in terms of {\tt SUM} and {\tt COUNT} in order to be able
to derive {\tt AVG} using the materialized view. Our system 
supports the standard set of aggregates: {\tt COUNT}, {\tt SUM}, 
{\tt AVG}, {\tt MIN}, and {\tt MAX}. Our implementation of 
{\tt rewriteAgg} is rather straightforward, and thus we omit for brevity;
however, see Srivastava et al.~\cite{mav2} for details.
If an aggregate exists, this requires adding the \cfree\ column to the {\tt GROUP} list of \qbe{},
which is done in lines 9--10.
Note that we do not
include terms in the {\tt ORDER} and the {\tt LIMIT} clauses during
materialized view creation.

For some queries and columns, delaying filters might lead to 
materialized view definitions that do not suit our split-execution 
scenario due to physical limitations on the backend or the
frontend. The materialized view query can exhaust the resources of
the backend, for example, when the size of the intermediate output
exhausts the physical resources of the backend. The size of the final
materialized view may also not fit into memory available
at the frontend. These two considerations must be addressed
when deciding which columns to free. 

Our current implementation depends on a set of rules
for deciding which columns can be freed. We disallow 
freeing columns that interact with other columns because they are
likely to exhaust physical resources. Namely, a column cannot be in a
condition on multiple columns such as a join condition or a complex
predicate involving other columns. We also disallow freeing a column
mentioned in an {\tt OR} clause. Finally, our
system allows for subqueries but disallows freeing a column referenced
in the subquery.

\subsection{View Query}
\label{csql}

At query time, Afterburner must validate whether a materialized view
can be used to answer a query.  This is accomplished using rules
that work on a parse tree representation of an SQL query, which can
be evaluated efficiently (negligible time in our
evaluations). Since we expect only a small number of materialized
views at the frontend, and that the data scientist is issuing
queries in rapid succession, we believe that a simple rule-based
approach is sufficient.

As demonstrated in our API, our task is query validation against
a single materialized view.
An important result from Larson and Zhou~\cite{larsonspgj} is that a
set of rules can be used to verify query coverage for the query class
SPJOG (Select, Project, Join, and Outer join queries with a possible
Group by). Thus, for a new query (\qnew), a query optimizer should be
able to rewrite it against the local materialized view (\qbe)
transparently.  In our prototype, we use simpler conditions to validate
the query against a materialized view because our goal is to
accelerate queries at the frontend---as 
opposed to general-purpose query optimization.

Our set of 
rules is simpler because of two main reasons. First, our prototype
does not consider partial matching plans, which are plans that involve
both the materialized view and base relations. While mixed 
plans can potentially improve overall query latency, we do
not consider them in this setup since we do not assume access to 
the base tables at the client. Second, as we have mentioned in the 
previous section, we do not allow for predicates on the \cfree\ 
column in the backend query. This simplifies our frontend plans 
to plans involving a single relation (which is \qbe) with exactly matching
predicates (other than predicates on \cfree). Thus, the only
differences allowed are in the {\tt GROUP BY} clause
and the {\tt WHERE} clause, and only involving the \cfree\ column.

\begin{myalgo}[t]
\begin{alltt}
\textnormal{
\textbf{Input: }\texttt{Query} \qnew, \texttt{Query} \qbe
\texttt{ 1: }{  }\qfe \(\leftarrow\) \textbf{new} \texttt{Query}()
\texttt{ 2: }{  }\qfe.\texttt{SELECT} \(\leftarrow\) \qnew.\texttt{SELECT}
\texttt{ 3: }{  }\textbf{for} c \(\in\) \qfe.\texttt{SELECT} \(\land\) \texttt{isAggregate}(c):
\texttt{ 4: }{      }c \(\leftarrow\) \texttt{deriveAgg}(c)
\texttt{ 5: }{  }\qfe.\texttt{FROM} \(\leftarrow\) \texttt{\{}\qbe.name\texttt{\}}
\texttt{ 6: }{  }\textbf{for} p \(\in\) \qnew.\texttt{WHERE} \(\land\) p \(\in\) \texttt{\{cfree\}}:
\texttt{ 7: }{      }\qfe.\texttt{WHERE} \(\leftarrow\) \qfe.\texttt{WHERE} \(\cup\) \texttt{\{}p\texttt{\}}
\texttt{ 8: }{  }\qfe.\texttt{GROUP} \(\leftarrow\) \qnew.\texttt{GROUP}
\texttt{ 9: }{  }\qfe.\texttt{ORDER} \(\leftarrow\) \qnew.\texttt{ORDER}
\texttt{10: }{  }\qfe.\texttt{LIMIT} \(\leftarrow\) \qnew.\texttt{LIMIT}
\texttt{11: }{  }\textbf{return} \qfe}
\end{alltt}
\caption{Generate VQ}
\label{algfe}
\end{myalgo}

\smallskip \noindent {\bf Materialized view matching conditions.}
When a new query is issued at the frontend with a
corresponding materialized view, Afterburner uses the following conditions
to validate whether \qbe\ can be used to answer (matches) \qnew:

\smallskip
\begin{small}
\begin{alltt}
{  cond 1: }\qnew.\texttt{SELECT} \(\subseteq\) \qbe.\texttt{SELECT}
{  cond 2: }\qnew.\texttt{FROM} \(=\) \qbe.\texttt{FROM}
{  cond 3: }\qnew.\texttt{JOIN} \(=\) \qbe.\texttt{JOIN}
{  cond 4: }\qnew.\texttt{WHERE} \(\setminus\) \texttt{\{}cfree\texttt{\}} \(=\) \qbe.\texttt{WHERE}
{  cond 5: }\qnew.\texttt{GROUP} \(\setminus\) \texttt{\{}cfree\texttt{\}} \(\subseteq\) \qbe.\texttt{GROUP}
\end{alltt}
\end{small}
\smallskip

\noindent Condition 1 checks that the {\tt SELECT} list of \qnew\ is a
subset of the \qbe.{\tt SELECT}, i.e., the new query is not
allowed to select new columns. Conditions 2 and 3 verify that the
{\tt FROM} and {\tt JOIN} lists of \qnew\ matches exactly \qbe.
Condition 4 checks that predicates in the {\tt WHERE} list of \qnew
match exactly the predicates in \qbe and may only add
predicates on the \cfree\ column. Condition 5 checks that all the 
{\tt GROUP} lists of \qnew (if any) are either already included
in the {\tt GROUP} list of \qbe or is the \cfree\ column. 

\smallskip \noindent {\bf View query generation.}
To generate the view query we use Algorithm~\ref{algfe}, which 
rewrites the new query (\qnew) against the materialized view (\qbe).
Line 1 initializes an empty frontend query (\qfe).
Line 2 adds all the terms in the {\tt SELECT} list of \qnew
into \qfe. Lines 3--4 derive aggregate terms from the materialized
view using the {\tt deriveAgg} function, e.g., {\tt COUNT} is derived as a sum of counts;
once again, see Srivastava et al.~\cite{mav2} for details. Line 5 adds the
name of the materialized view (\qbe) in the {\tt FROM} list of \qfe. 
Lines 6--7 only add predicates on the \cfree\ column to the {\tt WHERE}
clause of \qfe. For example, \qfe for Q6 only includes the two predicates
{\tt l\_shipdate >= DATE \textquotesingle1994-01-01\textquotesingle} and 
{\tt l\_shipdate < DATE \textquotesingle1995-01-01\textquotesingle}.
Finally, lines 8--10 add all the terms in the {\tt GROUP}, 
{\tt ORDER}, and {\tt LIMIT} lists to \qfe.

\subsection{Additional Discussion}
\label{view:discussion}

For clarity, the above exposition describes how to free a single
column, although our algorithm generalizes to freeing multiple columns
in a straightforward manner. However, in practice 
(see Section~\ref{eval:split}), materialized views
for freeing multiple columns are quite large since the materialized
view query generates the Cartesian product of the columns---typically,
this is more data than the client can handle.

At a higher level, an obvious shortcoming of our approach is that
the data scientist needs to explicitly provide hints via the
\texttt{free} operator. Ideally, this should be automatic---the
system should infer the subject of exploration, pre-fetch queries in
the background while the user is idle, learn from interactions to
prioritize memory usage, and automatically invalidate local views if
the source data changes. Tackling these challenges, however, requires
solving a number of issues that are orthogonal to the focus of
this paper, for example, user modeling to identify and anticipate
``interesting'' columns. 
These are interesting future directions but
beyond the scope of this paper. In our usage scenario, the
materialized views are transient, bounded at most by the life of a
browser tab. We envision that ``invalidation'' and ``maintenance'' of views are
as simple as refreshing the browser tab. Even though the views may
be short-lived, our split-execution experiments in
Section~\ref{eval:split} show that the data scientist ``recoups''
the extra cost of the materialized views for interactive data
exploration rather quickly.

\section{Evaluation}
\label{sec:eval}

Experimental validation of our work is divided into three parts.
First, we conducted microbenchmarks to understand the performance
characteristics of the in-browser execution environment and how it
compares to native code execution. Second, we examined the performance
of Afterburner on end-to-end SQL analytics within the browser (the TPC-H
benchmark), comparing our system against MonetDB, a well-known column
store, and LegoBase, the state of the art in compiled queries for SQL.
Finally, we tied all the threads in this paper together and conducted an
evaluation of interactive SQL analytics performance, taking advantage
of Afterburner split execution. For these experiments, we compared our 
architecture with the alternative deployments discussed in
Section~\ref{section:background}.

\subsection{Microbenchmarks}
\label{sec:micro}

\begin{table}[t]
\center
\resizebox{0.38\textwidth}{!}{
\begin{tabular}[t]{|l|l|r|l|l|}
\hline
{\bf Type}	& {\bf Filter} 	    & {\bf Selectivity}	& {\bf Column Name} \\ \hline \hline
Integer 	&{\tt == 0}        & 1.0       	&{\tt o\_shippriority}  \\ \hline
Float   	&{\tt > 555.5}      & 1.0   	&{\tt o\_totalprice}   \\ \hline
String  	&{\tt == \textquotesingle 1-URGENT\textquotesingle}& 0.2    	&{\tt o\_orderpriority}\\ \hline
\end{tabular}
}
\vspace{0.2cm}
\caption{Filter predicates in our microbenchmarks.}
\label{tab:micro}
\end{table}

In our microbenchmarks, we focused on simple filter queries that
count the number of records matching a predicate on a particular
column. For this, we examined filtering integers,
floats, and strings. Table~\ref{tab:micro} shows the names of the
columns, along with the filters and their selectivities. For these
experiments, we scanned the \texttt{orders} table from TPC-H data at a
scale of 10 GB (15 million records). We examined the following
hand-written programs in JavaScript:

\begin{list}{\labelitemi}{\leftmargin=1em}

\item JavaScript without asm.js or typed arrays, which we refer to as
  J1 for convenience.

\item JavaScript without asm.js but with typed arrays, which we refer
  to as J2 for convenience.

\item JavaScript with both asm.js and typed arrays, which we refer to
  as J3. This takes advantage of all JavaScript optimizations that we
  described in Section~\ref{section:design}.

\end{list}

\noindent 
Note that the option of evaluating asm.js without typed arrays
is not possible since the asm.js compiler disallows accessing
JavaScript objects.
We compared the above conditions against hand-written C++ programs,
compiled under the following conditions:

\begin{list}{\labelitemi}{\leftmargin=1em}

\item GCC (g++ 5.4.0) using optimization level \texttt{-O3}, which we
  refer to as GCC-O3 for convenience.

\item GCC, but without any optimizations.

\item Clang (v 3.9.0), with optimization level \texttt{-O3}, which we
  refer to as Clang-O3 for convenience.

\item Clang, but without any optimizations.

\end{list}

\noindent Experiments were run on a desktop with a 2.7~GHz Intel i5-5250U
processor (4 cores, 3~MB of cache) and 8~GB of RAM, running Ubuntu
16. The JavaScript conditions ran in Mozilla Firefox (v50).  We
ran each condition five times for warmup and report the average
runtime over the next five trials. Note that warm runs allow the JavaScript
conditions to benefit from JIT code caching.

\begin{table}[t]
\center
\resizebox{0.38\textwidth}{!}{
\begin{tabular}[t]{|l||rr|rr|rr|}
\hline
\frccdo{\bf Condition}&\multicolumn{2}{c|}{{\bf Integer}} &\multicolumn{2}{c|}{{\bf Float}}&\multicolumn{2}{c|}{{\bf String}} \\ \hline \hline
GCC-O3  &4.4 &             &4.4 &              &56.4 &\\ \hline
GCC     &37.0&(8.4$\times$)&47.8&(10.9$\times$)&103.8&(1.8$\times$)\\ \hline
\hline
J3      &16.9&(3.8$\times$)&22.7&(5.2$\times$) &60.8 &(1.1$\times$)\\ \hline
J2      &20.1&(4.6$\times$)&25.7&(5.8$\times$) &94.7 &(1.7$\times$)\\ \hline
J1      &28.6&(6.5$\times$)&28.5&(6.5$\times$) &406.1&(7.2$\times$)\\ \hline
\hline
Clang-O3&3.8 &             &4.0 &&60.4&\\ \hline
Clang   &40.4&             &45.4&&97.0&\\ \hline
\end{tabular}
}
\vspace{0.2cm}
\caption{Microbenchmarks showing query latencies (ms).}
\label{tab:microlat}
\end{table} 

Results are shown in Table~\ref{tab:microlat}, where we report query
latencies in milliseconds and slowdown with respect to GCC-O3 in
parentheses. 
For these experiments we exclude compilation times for C++ and asm.js.
Comparing GCC-O3 with fully-optimized JavaScript (J3), we
see a slowdown of roughly 4--5$\times$ for filter predicates
over integers and floats, but only slowdown of 10\% for
strings. Detailed analyses show that the performance advantage of
GCC-O3 comes from automatic loop unrolling and the generation of SIMD
instructions. On the other hand, strings are only marginally slower
with J3 because of the extra level of indirection associated with a
standard \texttt{(char **)} representation of strings in C++, where
the query latency is dominated by memory latencies associated with pointer dereferencing.
We see the effectiveness of SIMD instructions and loop unrolling with
GCC (without any optimizations), which is actually slower than
fully-optimized JavaScript. Comparing J3 with J2 (typed arrays but no
asm.js) and J1 (no optimizations), it is clear that both features of
JavaScript contribute to performance and in a cumulative fashion.
Finally, our results show that Clang performance is on par with GCC
performance. This provides a sanity check as one of the
conditions in the next section uses Clang.

In Table~\ref{fig:pmu}, we show the output of CPU performance
counters comparing GCC-O3 and J3, using the \texttt{perf-stat} profiling tool.
Measuring such low-level performance for GCC-O3 is straightforward since the
code runs as a separate process. On the other hand, measuring
CPU performance for J3 running inside the browser is
challenging because J3 runs as a thread inside the browser's main
process.
Thus, even in an idle state, without user interactions,
the browser process is active (e.g., running event loops)
and generating data captured as part of profiling. In order to 
minimize such interference in our measurements we took the following
steps:\ First, we repeated each run 500 times and report the average.
Second, we minimized the size of the browser window,
leaving only the JavaScript console available to run our benchmark scripts. This helps
to avoid spurious interactions.

\begin{table}[t]
\center
\resizebox{.48\textwidth}{!}{
\begin{tabular}{|l||r|r||r|r||r|r||r|r|}
\hline
                      &\multicolumn{2}{c||}{\bf Branches} & \multicolumn{2}{c||}{\bf Mispredicts} & \multicolumn{2}{c||}{\bf D1 misses} & \multicolumn{2}{c|}{\bf I1 misses} \\ \hline \hline
\frccdo{\bf Type}&\frcc{\bf O3}&\frccd{\bf J3}&\frcc{\bf O3}&\frccd{\bf J3}&\frcc{\bf O3}&\frccd{\bf J3}&\frcc{\bf O3}&\frcc{\bf J3} \\ \hline
Integer  &  3.7m & 45m &   37 & 7.6k &  940k & 970k & 640 & 12k \\ \hline
Float    &  3.8m & 45m &   26 & 3.4k &  940k & 950k & 520 & 12k\\ \hline
String   & 33.0m & 51m & 3.2m & 3.3m &  3.6m & 3.6m & 7700 & 21k \\ \hline
\end{tabular}
}
\vspace{0.2cm}
\caption{Performance counters comparing GCC-O3 vs.\ J3.}
\label{fig:pmu}
\end{table}

Examining the number of branches:\ for GCC-O3, which uses SIMD
instructions and is able to process four integers or floats at a time, we
observe an average of $0.25$ branches per record, as expected (in these
microbenchmarks we are scanning 15 million records). In comparison, J3
averages three branches per record, which explains the latency gap. Note
that using SIMD is a compiler optimization, and so it is
possible that future JavaScript runtimes will be smart enough to take
advantage of SIMD instructions also. Due to loop unrolling for
integers and floats, there are barely any branch mispredicts for
GCC-O3. In terms of data cache misses for integers and floats, we
observe roughly the same counts, since typed arrays are implemented
using native arrays. For instruction cache misses, we see lower counts
for GCC-O3 since SIMD code is more compact.
Note that, however, when it comes to strings, the SIMD instructions
are no longer applicable, and we observe similar counts for branch
mispredicts (although GCC-O3 generates code with fewer branches).

In summary, these microbenchmarks show that the JavaScript execution
environment inside modern browsers is very efficient, in some cases
rivaling native performance. This is not surprising:\
the prevalence of complex JavaScript applications (e.g., Gmail, Facebook,
etc.)\ in running modern websites means that countless hours have been
devoted to optimizing JavaScript performance---we are the
beneficiaries of all this effort.
Native code is still much faster in our
microbenchmarks because of SIMD instructions and loop unrolling, but
these optimizations do not appear to be possible on more complex queries:\
this is confirmed by experimental results in Section~\ref{eval:tpc},
where modern compiled query techniques running natively do not outperform
Afterburner by anywhere close to the margins reported in the microbenchmarks.
Overall, the impression that ``JavaScript is slow'' is
simply a myth---high-performance SQL analytics engines in JavaScript are possible. We turn
to this next.

\begin{figure*}[t]
  \centering
  \begin{minipage}{.7\textwidth}
  \includegraphics[width=\textwidth, keepaspectratio]{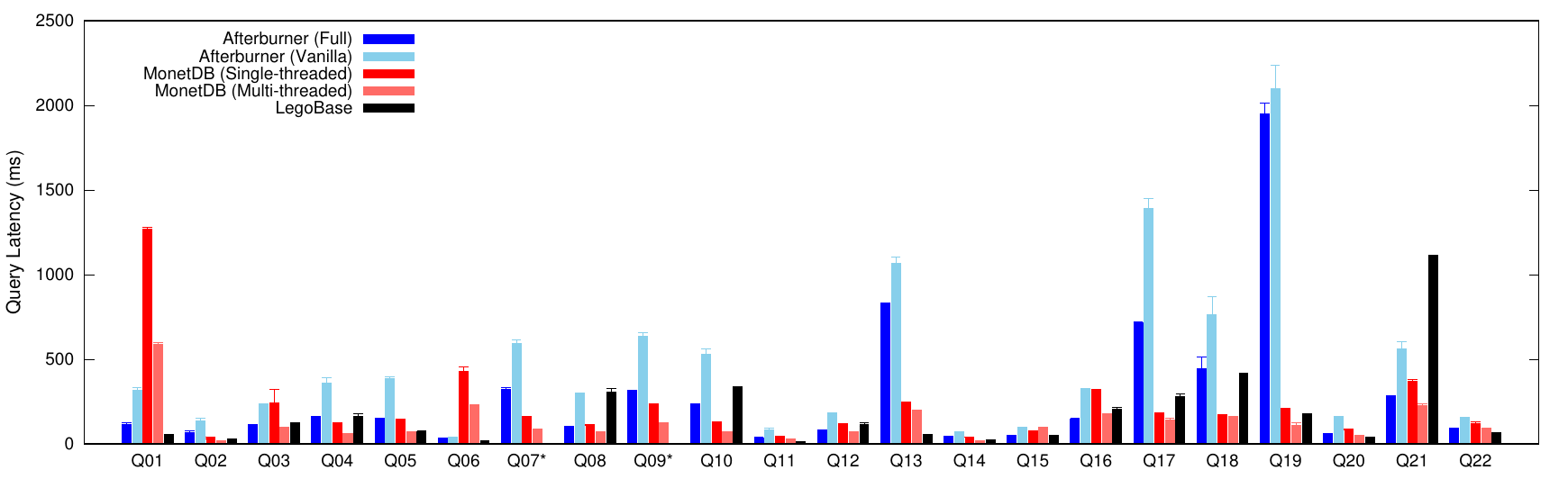}

  \end{minipage}
  \begin{minipage}{.29\textwidth}
  \includegraphics[width=\textwidth, keepaspectratio]{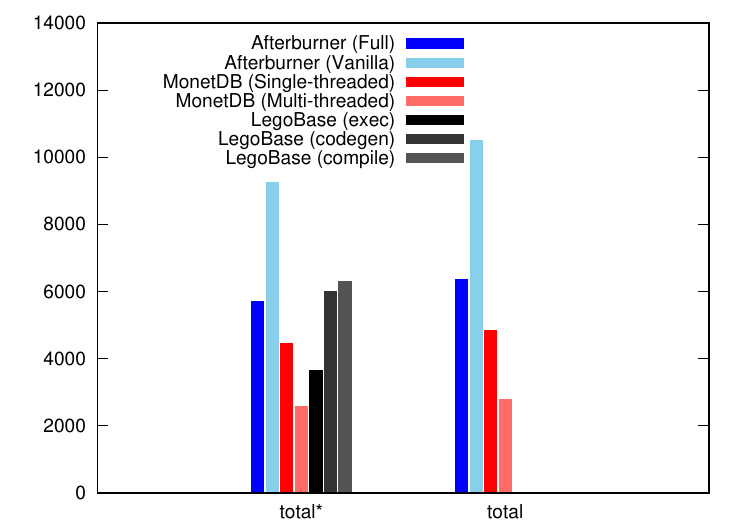}

  \end{minipage}
  \caption{Query latency in ms for TPC-H queries, comparing
      Afterburner with MonetDB and LegoBase:\ query latencies are over
      five trials with 95\% confidence intervals on the left and total
      query latency on the right.}
  \label{fig:allq}
\end{figure*}

\subsection{In-Browser Analytics Performance}
\label{eval:tpc}

We are, of course, not the first to explore data management
inside the browser. A detailed comparison with SQL-in-JavaScript
solutions is presented in Appendix~\ref{sec:existing}, but the
high-level finding is that current JavaScript SQL engines such
as Lovefield and Sql.js are not able to efficiently support
analytical queries for even modestly-sized databases. The
performance and scalability of Afterburner is far superior to either
of these systems.

To examine the end-to-end query performance of our
approach, we compared the full Afterburner system against the
following:

\smallskip \noindent {\bf MonetDB} (v11.23.13) is an open-source
analytical RDBMS that takes advantage of columnar storage and
vectorized execution. We used the TPC-H test harness written by the
developers of the system. We ran MonetDB under two
conditions:\ For a fair comparison, MonetDB was configured to use a
single thread, since code running inside a browser tab is single
threaded. In addition, we also evaluated MonetDB with all cores
enabled as another point of reference.

\smallskip \noindent {\bf LegoBase}~\cite{lego,dblab}\footnote{The
term LegoBase is a bit ambiguous since it refers to several
different software components, but in this context we specifically
refer to the authors' SIGMOD 2016 paper that we replicate and
compare against.} is an open-source in-memory RDBMS that represents
the state of the art in code generation. LegoBase takes as input a
representation of the physical plan of a query and then generates code for
this plan. LegoBase uses its knowledge of the target programming
language and database statistics to pick the best physical operators
for a query. In our experiments, we
configured LegoBase to target C, to
generate code under what the authors call the ``compliant'' condition,
which generates query operators that are compliant with the TPC-H
benchmark. We measured the time to generate the target C code ({\tt
  codegen}), time to compile the target code using Clang ({\tt
  compile}), and query execution latency (which {\it does not} include
{\tt codegen} and {\tt compile} stages).  For all experiments we used
available open-source code\footnote{\url https://github.com/epfldata/dblab}
and sought guidance from the authors of the papers to ensure that
we were using their system properly. Our runs are generally consistent
with the results reported in their papers.

\smallskip \noindent {\bf Afterburner (Vanilla).} To isolate the
impact of asm.js in the performance of Afterburner, we also evaluated
Afterburner (Vanilla), a variant that uses typed arrays but not asm.js (i.e., the
compiled query plans were evaluated with asm.js optimizations
disabled). Thus, Afterburner (Vanilla) resembles the J2 condition from
the previous section.

\smallskip \noindent Thus, we compared five different
experimental conditions in total:\ Afterburner (Full), Afterburner (Vanilla), MonetDB
(Single-threaded), MonetDB (Multi-threaded), and LegoBase.  We have
specific rationale for each of these comparisons:\ MonetDB is a mature
and stable implementation of well-known techniques for analytical data
processing (albeit the techniques are around a decade old). On the
other hand, LegoBase represents the ``latest and greatest'' research
on compiled queries. It is best described as a research toolkit
comprised of many different pieces, as opposed to a complete
RDBMS. Finally, Afterburner (Vanilla), compared to the full system,
allows us to separately study the contributions of typed arrays and
asm.js to overall performance.

In these experiments, we used the TPC-H benchmark at a scale
factor of 1 GB, which corresponds to 6 million records in the
\texttt{lineitem} table. This represents a rough upper bound on the
amount of data that a commodity desktop or laptop can comfortably
hold today. Experiments were performed on the same client machine as
in the previous section.

All our measurements were on a warm cache---we first ran each query
five times, and then took measurements over five trials. For LegoBase,
we warmed up codegen and query compilation in the same way. For
Afterburner (both the full and vanilla configurations), measured
latency includes query compilation overhead and all data are
explicitly loaded in memory. For MonetDB, we confirm that all data are
cached in the underlying OS buffer caches.

Figure~\ref{fig:allq} (left) shows the query latency of all 22 TPC-H
queries for all our experimental conditions; detailed
running times are shown in Appendix~\ref{appendix:detail}. As expected,
multi-threaded MonetDB is faster than single-threaded MonetDB (by
1.7$\times$), but this is not a fair comparison since one condition has access to
more hardware resources. Leaving aside MonetDB (Multi-threaded),
across all queries, MonetDB (Single-threaded) is the fastest for 6
queries, LegoBase is the fastest for 11 queries, and Afterburner (Full) is
the fastest for 5 queries. MonetDB (Single-threaded) beats Afterburner
(Full) for 11 queries and LegoBase beats Afterburner (Full) for 11
queries. From the right plot in Figure~\ref{fig:allq}, MonetDB
(Single-threaded) completed all 22 queries in 4.8s, compared to 6.4s
for Afterburner (Full), which makes Afterburner (Full) 1.3$\times$
slower. LegoBase runs out of available physical memory on our client
desktop for Q7 and Q9, so we computed total* as the running time
over just the remaining 20 queries:\ LegoBase finishes all of them
in 3.7s, MonetDB (Single-threaded) finishes in 4.5s, while Afterburner
(Full) finishes in 5.7s, making our system 1.5$\times$ and 1.3$\times$
slower, respectively. For reference MonetDB (Multi-threaded)
completes all 22 queries in 2.8s and the 20 queries in 2.6s.

The comparison between Afterburner (Full) and LegoBase gives us a
sense of how our approach compares to modern compiled query
techniques. Note that all measurements of our system represent
end-to-end execution time, while latencies for LegoBase do not
include code generation or compilation. The total time to generate
code and to compile the TPC-H queries for the 20 queries
under the total* condition was 7.2s and 7.4s, respectively.
To be fair, LegoBase was
not optimized for interactive query exploration---both the codegen
and compilation stages {\it individually} are longer than end-to-end
execution for Afterburner (Full). LegoBase is perhaps more suitable for
even larger data conditions and repeated execution, where the time
spent in codegen and compilation can be amortized over longer query
execution times.

Comparing Afterburner (Full) with Afterburner (Vanilla) allows us to
tease apart the performance contributions of Afterburner's two different
JavaScript optimizations:\ typed arrays and asm.js. Indeed,
disabling asm.js optimizations yields a total latency of 10.5s on all 22 queries
(1.6$\times$ slower). To assess the impact of typed arrays, we tried
loading the TPC-H data into the browser as JavaScript objects, and
unsurprisingly, it failed (non-responsive browser,
out of physical memory). Overall, these results are consistent with the microbenchmarks
in Section~\ref{sec:micro} and show that both optimizations of Afterburner
are critical to performance and scalability. Typed arrays enable
compact, in-memory representations, while asm.js enables efficient
code execution.

In summary, these experiments show that it is possible to
build a high-performance analytical RDBMS in JavaScript. While its
performance still lags MonetDB and LegoBase, both of which run
natively, we find Afterburner's performance quite impressive,
considering that it runs completely in the browser!

\begin{table*}[t]
\centering\begin{small}
\begin{tabular}{|l|l||r||r|r|r||r|r|r||r|}
\hline
&\frccd{Free column}    & {\bf A}  & MVQ & MV size & MV copy & {\bf B} & {\bf C} & {\bf D} & Breakeven \\
&\frccd{(cardinality)}  & (ms)     &  (ms) & (records) & (ms)  &  (ms) & (ms) & (ms) & (D vs.\ A)\\ \hline
\hline
Q1a & l\_shipdate (2526)    & 34,330 & 119,843 &3,817&277 &19&16&13 & 4 \\ \hline
\hline
Q2a & p\_size (50)          &  2,556 &  27,047 &2,365,583&155,826 &82&28&43 & 72 \\ \hline
Q2b & p\_type (150)         &  2,556 &   4,405 &236,211&16,380 &331&128&44 & 9 \\ \hline
\hline
Q3a & c\_mktsegment (5)     & 17,170 &  19,051 &5,662,337&107,526 &198&132&1,147 & 8 \\ \hline
Q3b & o\_orderdate (2406)   & 17,170 &  36,142 &16,553,365&268,052 &233&174&1,280 & 18 \\ \hline
\hline
Q4a & o\_orderdate (2526)   &  6,924 & 119,843 &12,030&277 &11&9&12 & 18 \\ \hline
\hline
Q5a & r\_name (5)           &  7,712 &   8,890 &25&203 &14&10&11 & 2 \\ \hline
Q5b & o\_orderdate (2406)   &  7,712 &  10,125 &12,030&517 &10&8&13 & 2 \\ \hline
\hline
Q6a & l\_shipdate (2526)    &  4,362 &   9,815 &2,526&247 &13&12&7 & 3 \\ \hline
Q6b & l\_discount (11)      &  4,362 &   4,590 &11&375 &12&9&6 & 2 \\ \hline
Q6c & l\_quantity (50)      &  4,362 &   8,100 &50&420 &11&10&6 & 2 \\ \hline
\hline
Q7a & l\_shipdate (2526)    &  8,388 &   8,390 &5,052&649 &11&10&11 & 2 \\ \hline
\hline
Q12a& l\_shipmode (7)       &  5,420 &   5,760 &7&28 &12&11&13 & 2 \\ \hline
Q12b& l\_receiptdate (2554) &  5,420 &   7,219 &4,985&163 &10&13&12 & 2 \\ \hline
\hline
Q14a& l\_shipdate (2526)    &  4,052 &  29,298 &2,526&364 &9&10&7 & 8 \\ \hline
\hline
Q17a& p\_brand (25)         & 20,374 & 142,773 &25&468 &14&11&6 & 8 \\ \hline
Q17b& p\_container (40)     & 20,374 & 120,131 &40&678 &9&11&7 & 6 \\ \hline
\hline
Q20a& n\_name (25)          &  8,546 &  10,000 &447,508&10,699 &565&243&22 & 3 \\ \hline
Q21a& o\_orderstatus (3)    & 23,540 &  30,202 &96,037&2,692 &68&21&66 & 2 \\ \hline
Q21b& n\_name (25)          & 23,540 &  76,754 &999,953&16,221 &129&67&94 & 4 \\ \hline
\end{tabular}
\end{small}
\vspace{0.2cm}
\caption{Experimental results comparing Deployments A, B, C, and D on
  interactive SQL data exploration scenarios derived from the TPC-H
  benchmark. Each variant query describes a different ``freed'' column.}
\label{tab:mavdet}
\end{table*}

\subsection{Split-Execution Performance}
\label{eval:split}

Our final set of experiments brings together all of the threads
discussed in this paper to evaluate the performance of interactive SQL
analytics in Afterburner, taking advantage of our split-execution
techniques. Referring back to Section~\ref{section:background}, we
compare our proposed Deployment D (Afterburner) against the
alternatives A, B, and C.

In order to create a query mix that captures typical interactive SQL
analytics tasks, we once again draw inspiration from the TPC-H
benchmarks. For evaluation, we considered all the TPC-H queries with
columns that can be ``freed'' based on our approach described in
Section~\ref{section:split}. For each TPC-H query, we generated
variant queries as appropriate:\ for example, for Q6, we can free the
columns \texttt{l\_shipdate}, \texttt{l\_discount},
\texttt{l\_quantity}, which yields Q6a, Q6b, and Q6c, respectively.
In total, 12 queries from the TPC-H benchmark are amenable to this
treatment. For each query, different columns can be freed. In some cases, the
materialized views were too large to store at the client side; these
queries were discarded in our evaluation. 
In total, we created 20 variant queries,
shown in the first column of Table~\ref{tab:mavdet}; the second column
names the column in the query that is freed (e.g.,
\texttt{l\_shipdate}) and its cardinality. 
As discussed in Section~\ref{view:discussion}, Afterburner
is able to free more than one column at a time, but this is often
not practical due to the size of the materialized views.
Thus, in our evaluation, we only considered cases
where one column is freed at a time.

Our experiments used a data warehouse as defined by TPC-H
at a scale factor of 100 GB, which yields a \texttt{lineitem} table
with 600 million records. The backend in all scenarios is MonetDB running
on a server with dual 8-core Intel Xeon E5-2670 processors (2.6~GHz)
with 256~GB of memory on Ubuntu 14.04, configured to take advantage of
all available hardware resources. The frontend machine in Deployments
C and D is the same as the desktop describe in the previous
sections. In Deployment C, we run MonetDB on the local client machine,
and in Deployment D, we run Afterburner in the browser. As a minor
detail, in both Deployment B and Deployment C we ran MonetDB with only
a single core because the sizes of the materialized views are
sufficiently small that single core performance is actually better
than multi-core performance---the overhead associated with multi-core
query execution is more than the performance gained via
parallelism. For Deployment C, latency is measured from the perspective
of the browser, i.e., materializing the result set inside the browser.
In all our experiments, the client and backend server are
both located in the same building, with round trip ping times less than
half a millisecond.  All performance measurements reported below were
on a warm cache and we report averages over five trials.

Results of our experiments are shown in Table~\ref{tab:mavdet}. The
column marked ``A'' shows the query latencies under Deployment A,
which does not take advantage of materialized views. The ``free''
clause does not apply; the query latency is simply the query execution
time of the original TPC-H query posed against the backend. For ease
of comparison, we simply repeat the latency in the row for each
query variant. Deployments B, C, and D all take advantage of
materialized views:\ the latency of the query that generates this
materialized view is reported under the column ``MVQ''. The size of
the materialized view (in number of records) is reported under the column
``MV size'', and the time it takes to copy the materialized view from
the backend to the client is reported in the column ``MV copy''
(applicable only for Deployments C and D). The latency of the view
queries under Deployments B, C, and D are the columns marked
``B'', ``C'', and ``D'', respectively.

Continuing with our running example of Q6 from
Section~\ref{section:background}:\ from the results table we see that the
unmodified query takes around 4.4s. If we wish to free the
\texttt{l\_shipdate} column, which has a cardinality of 2526, the
materialized view query takes around 9.8s. However, all subsequent
queries that involve only changes to predicates on {\tt l\_shipdate}
only take $\sim$10ms to run against the materialized view using any of the
Deployments B, C, and D. Of course, in the case of Deployments C and
D, the materialized view needs to be copied over to the client
machine, which takes 247ms.

The performance of the view queries for Q6 (comparing
Deployments B, C, and D) are comparable, but
performance aside, these deployments manifest all the tradeoffs
discussed in Section~\ref{section:background}. The performance of
Deployment B is affected by backend query load (e.g., concurrent
queries by multiple data scientists) as well as variability in network
latencies. In our case the backend is less than half a millisecond
away, but in an enterprise cloud deployment, latencies could be much
longer. For example, the round trip ping time between our client and an
arbitrary instance on Amazon's EC2 service running in the US East
region is around 25ms. To be fair, though, link latencies will also
affect the time it takes to copy the materialize view over to the
client (in the case of Deployments C and D), but the advantages of
local interactions remain in eliminating all subsequent need for
interactions with the backend.

Comparing Deployments C and D, with Afterburner (Deployment D), we
have eliminated the need to have a local RDBMS installation---complete
with all the headaches of maintaining a local software stack. With
Afterburner, we achieve breakeven in three queries:\ if the data scientist
issues three queries using the same query template as part of
interactive data exploration, we make up for the fact that the initial
materialized view query takes longer (even after accounting for the
cost of transferring the materialized view). This is shown in the
final column in Table~\ref{tab:mavdet}.

Looking at results across all queries in Table~\ref{tab:mavdet}, we
see that in most cases Afterburner (Deployment D) achieves performance
parity with MonetDB in both Deployments B and C. Where there are
performance differences, they are for the most part negligible. However,
there are a few special cases to note:\ for Q3, Afterburner queries
take substantially longer to execute in the browser. In this case, the
materialized view is quite large and the query involves a top-$k$.
MonetDB is able to optimize this into a scan, whereas Afterburner
inefficiently sorts all records before taking the top $k$. The
performance difference, in this case, is caused by deficiencies in query
optimization in our implementation, not an inherent limitation of our approach.

Other results worth discussing are performance differences
between Deployments B and C for some queries---since they are both
running MonetDB and the differences are much larger than can be
explained solely by hardware (the backend vs.\ the client). This is
most evident for Q20a, where the view query for Deployment B takes
more than twice as long as the view query for Deployment C. We
attribute these differences to the cost of transferring the result set
over:\ in this case, the result set contains 18k records. The latency in
Deployment C includes the overhead of loading the result set
into the browser.

To summarize, it would be fair to characterize Deployment D as
achieving performance parity with Deployment C. This is consistent
with results from the previous section, where we examined end-to-end
SQL analytics within the browser in comparison to
MonetDB. Therefore, with Afterburner we can eliminate the need for
client-side deployment of an RDBMS (with all its associated
administrative and maintenance headaches) without compromising
performance.

\section{Related Work}

\smallskip \noindent {\bf Split execution.}
The idea of splitting query execution across different layers of the stack
is of course not new~\cite{Franklin_etal_SIGMOD1996,semcache,DBLP:conf/icde/BowmanS07}.
This paper revisits the same idea for accelerating interactive analytical
SQL queries given two new conditions:\ First, the ability to execute
analytical queries on the frontend in a JavaScript environment.
Second, providing the data scientist an
easy way to define the target of exploration around a query template.
Our idea of ``freeing'' columns is related to the work
of Koudas et al.~\cite{Koudas:2006:RJS:1182635.1164146} in ``relaxing''
join and selection queries, but their goal is to ``back off'' from 
queries that return empty results. Also related is the semantic 
pre-fetching idea of Bowman and Salem~\cite{DBLP:conf/icde/BowmanS07},
who try to predict future queries based on past history; in our case, 
we require explicit hints from the user.

\smallskip \noindent {\bf Compiled queries.}
The compiled query approach of Afterburner takes after systems like
HIQUE~\cite{hiq}, LegoBase~\cite{lego,dblab}, Proteus~\cite{proteus}, and
HyPer~\cite{Neumann11}, which have recently popularized
code generation for relational query processing. With the exception
of targeting JavaScript and dealing with all the limitations associated
with running inside a browser, our query
compilation techniques are fairly standard. We use a template-matching
approach to generate compiled queries; for example, simple \texttt{SELECT}-\texttt{WHERE} queries
are converted into {\tt for} loops over the appropriate ranges of the typed
array holding the data. In our current prototype, there is no
query optimization to speak of, as we only have a single hash-based physical plan
for both joins and group bys.

One well-known drawback of query compilation is that compiling generated code using a
tool like {\tt gcc} can overshadow its benefits for short-running
queries. Much research has gone into alleviating this issue, such as the use of
an intermediate representation like LLVM~\cite{Neumann11}.
In our work, however, we have found
compilation overhead to be negligible, primarily because compilation
speed is already something browsers optimize for since all
JavaScript code on the web is stored as text.

\smallskip \noindent {\bf Materialized views.}
Taking advantage of materialized views to optimize complex queries is
a well-studied problem dating back decades~\cite{mav,mav2,ast,ms01}.
There are, however, substantial differences with our approach:\ in
most previous work, materialized views are long-lived and
carefully-considered by a database administrator, not
built willy-nilly on an ad hoc and per-query basis---which is the
approach that we take. For us, materialized views are transient and
lightweight, precisely because they are shipped over to the browser
and can be discarded when done. 
With our in-browser JavaScript engine, not only is the integration
seamless, but subsequent interactions can happen
without the backend.

\smallskip \noindent {\bf Physical design tuning.}
In a sense, Afterburner shares similarities with physical design 
advisers~\cite{dbtune1,dbtune2,dbtune3} since that Afterburner picks
materialized views automatically, without requiring special expertise
or adding an extra administrative burden. Physical design advisers,
choose a set of materialized views that optimize a query workload under
constraints, such as fitting a storage budget. The query optimizer is
heavily engaged in the process of identifying a candidates space to
pick from (i.e. depend on the query planner enumeration 
space~\cite{dbtune3} to look for possible materialized views). 
In Afterburner, we apply query re-writing rules to generate the 
materialized view definition which can generate materialized views
that would not be considered by a query planner.

\smallskip \noindent {\bf Shared work and shared data.}  Our work also
has similarities with techniques targeted at accelerating streams
of queries by sharing parts of their plans~\cite{qpipe,shareddb2}.
Similar to this line of work, Afterburner optimizes multiple queries by
finding a common subplan (the materialized view). However, the
optimization goals and approaches are very different. Our goal is
to minimize the latency of a family of queries anchored around a
single query template for a specific user, while subplan sharing
techniques have the goal of improving overall system performance,
e.g., higher query throughput, typically across multiple users. Once
again, despite superficial similarities, Afterburner targets a
completely different point in the design space.

\smallskip \noindent 
{\bf Approximate query answering.}
Related to materialized views are
techniques that provide faster but approximate answers to 
user queries~\cite{aqua,blinkdb,Agarwal:2012:BDI:2367502.2367533,verdict,alexg}. 
One obvious difference is that Afterburner delivers exact answers.
More importantly, though, as already discussed in Section~\ref{section:background},
this thread of work provides additional support and motivation for the usefulness of
our query scenario. At a high level, all the approaches cited above depend
on a sequence of related queries that allow the system to approximate
the distribution of the underlying data. Thus, these papers
share similar intuitions as our own work but exploit them in very
different ways.

\section{Conclusion}

This paper explores the somewhat unconventional idea of building data
management capabilities directly in the browser. We empirically show
that modern JavaScript runtimes are capable of supporting an
analytical RDBMS whose query performance rivals state-of-the-art
techniques running natively. Based on this, we propose a novel
split-execution strategy to support template-based interactive SQL
analytics. Experiments show that we achieve performance parity with
other deployment architectures, but with a simpler ``it's just a
webpage'' design. In other words, we can have our cake and eat it too!

%\bibliographystyle{abbrv}
%\bibliography{afterburner} 

\begin{thebibliography}{10}

\bibitem{Abadi_etal_SIGMOD2006}
D.~J. Abadi, S.~R. Madden, and M.~C. Ferreira.
\newblock Integrating compression and execution in column-oriented database
  systems.
\newblock {\em {SIGMOD}}, pages 671--682, 2006.

\bibitem{aqua}
S.~Acharya, P.~B. Gibbons, V.~Poosala, and S.~Ramaswamy.
\newblock The {Aqua} approximate query answering system.
\newblock {\em {SIGMOD}}, pages 574--576, 1999.

\bibitem{blinkdb}
S.~Agarwal, B.~Mozafari, A.~Panda, H.~Milner, S.~Madden, and I.~Stoica.
\newblock {BlinkDB}: Queries with bounded errors and bounded response times on
  very large data.
\newblock {\em {EuroSys}}, pages 29--42, 2013.

\bibitem{Agarwal:2012:BDI:2367502.2367533}
S.~Agarwal, A.~Panda, B.~Mozafari, A.~P. Iyer, S.~Madden, and I.~Stoica.
\newblock Blink and it's done: Interactive queries on very large data.
\newblock {\em {PVLDB}}, 5(12):1902--1905, 2012.

\bibitem{dbtune1}
S.~Agrawal, S.~Chaudhuri, and V.~R. Narasayya.
\newblock Automated selection of materialized views and indexes in {SQL}
  databases.
\newblock {\em {VLDB}}, pages 496--505, 2000.

\bibitem{dbtune5}
S.~Agrawal, V.~R. Narasayya, and B.~Yang.
\newblock Integrating vertical and horizontal partitioning into automated
  physical database design.
\newblock {\em {SIGMOD}}, pages 359--370, 2004.

\bibitem{Armbrust_etal_SIGMOD2015}
M.~Armbrust, R.~S. Xin, C.~Lian, Y.~Huai, D.~Liu, J.~K. Bradley, X.~Meng,
  T.~Kaftan, M.~J. Franklin, A.~Ghodsi, and M.~Zaharia.
\newblock {Spark} {SQL}: Relational data processing in {Spark}.
\newblock {\em {SIGMOD}}, pages 1383--1394, 2015.

\bibitem{DBLP:conf/icde/BowmanS07}
I.~T. Bowman and K.~Salem.
\newblock Semantic prefetching of correlated query sequences.
\newblock {\em ICDE}, pages 1284--1288, 2007.

\bibitem{dbtune2}
N.~Bruno and S.~Chaudhuri.
\newblock Automatic physical database tuning: {A} relaxation-based approach.
\newblock {\em {SIGMOD}}, pages 227--238, 2005.

\bibitem{semcache}
S.~Dar, M.~J. Franklin, B.~T. J{\'{o}}nsson, D.~Srivastava, and M.~Tan.
\newblock Semantic data caching and replacement.
\newblock {\em {VLDB}}, pages 330--341, 1996.

\bibitem{dbtune4}
K.~{El Gebaly} and A.~Aboulnaga.
\newblock Robustness in automatic physical database design.
\newblock {\em {EDBT}}, pages 145--156, 2008.

\bibitem{ElGebaly_Lin_SIGMOD2017}
K.~{El Gebaly} and J.~Lin.
\newblock In-browser interactive {SQL} analytics with afterburner.
\newblock {\em SIGMOD}, pages 1623--1626, 2017.

\bibitem{Franklin_etal_SIGMOD1996}
M.~J. Franklin, B.~T. {J\'{o}nsson}, and D.~Kossmann.
\newblock Performance tradeoffs for client-server query processing.
\newblock {\em {SIGMOD}}, pages 149--160, 1996.

\bibitem{alexg}
A.~Galakatos, A.~Crotty, E.~Zgraggen, C.~Binnig, and T.~Kraska.
\newblock Revisiting reuse for approximate query processing.
\newblock {\em {PVLDB}}, 10(10):1142--1153, 2017.

\bibitem{shareddb2}
G.~Giannikis, D.~Makreshanski, G.~Alonso, and D.~Kossmann.
\newblock Shared workload optimization.
\newblock {\em {PVLDB}}, 7(6):429--440, 2014.

\bibitem{ms01}
J.~Goldstein and P.-A. Larson.
\newblock Optimizing queries using materialized views: A practical, scalable
  solution.
\newblock {\em {SIGMOD}}, pages 331--342, 2001.

\bibitem{mav}
A.~Gupta, V.~Harinarayan, and D.~Quass.
\newblock Aggregate-query processing in data warehousing environments.
\newblock {\em {VLDB}}, pages 358--369, 1995.

\bibitem{qpipe}
S.~Harizopoulos, V.~Shkapenyuk, and A.~Ailamaki.
\newblock {QPipe}: {A} simultaneously pipelined relational query engine.
\newblock {\em {SIGMOD}}, pages 383--394, 2005.

\bibitem{proteus}
M.~Karpathiotakis, I.~Alagiannis, and A.~Ailamaki.
\newblock Fast queries over heterogeneous data through engine customization.
\newblock {\em {PVLDB}}, 9(12):972--983, 2016.

\bibitem{lego}
Y.~Klonatos, C.~Koch, T.~Rompf, and H.~Chafi.
\newblock Building efficient query engines in a high-level language.
\newblock {\em {PVLDB}}, 7(10):853--864, 2014.

\bibitem{Koudas:2006:RJS:1182635.1164146}
N.~Koudas, C.~Li, A.~K.~H. Tung, and R.~Vernica.
\newblock Relaxing join and selection queries.
\newblock {\em VLDB}, pages 199--210, 2006.

\bibitem{hiq}
K.~Krikellas, S.~Viglas, and M.~Cintra.
\newblock Generating code for holistic query evaluation.
\newblock {\em {ICDE}}, pages 613--624, 2010.

\bibitem{larsonspgj}
P.~Larson and J.~Zhou.
\newblock View matching for outer-join views.
\newblock {\em {VLDB}}, pages 445--456, 2005.

\bibitem{Neumann11}
T.~Neumann.
\newblock Efficiently compiling efficient query plans for modern hardware.
\newblock {\em {PVLDB}}, 4(9):539--550, 2011.

\bibitem{verdict}
Y.~Park, A.~S. Tajik, M.~J. Cafarella, and B.~Mozafari.
\newblock {Database Learning}: {T}oward a database that becomes smarter every
  time.
\newblock {\em {SIGMOD}}, pages 587--602, 2017.

\bibitem{Poess:2007:WYR:1325851.1325979}
M.~P{\"{o}}ss, R.~O. Nambiar, and D.~Walrath.
\newblock Why you should run {TPC-DS:} {A} workload analysis.
\newblock {\em {VLDB}}, pages 1138--1149, 2007.

\bibitem{dblab}
A.~Shaikhha, Y.~Klonatos, L.~Parreaux, L.~Brown, M.~Dashti, and C.~Koch.
\newblock How to architect a query compiler.
\newblock {\em {SIGMOD}}, pages 1907--1922, 2016.

\bibitem{mav2}
D.~Srivastava, S.~Dar, H.~V. Jagadish, and A.~Y. Levy.
\newblock Answering queries with aggregation using views.
\newblock {\em {VLDB}}, pages 318--329, 1996.

\bibitem{dbtune3}
G.~Valentin, M.~Zuliani, D.~C. Zilio, G.~M. Lohman, and A.~Skelley.
\newblock {DB2} {Advisor}: An optimizer smart enough to recommend its own
  indexes.
\newblock {\em {ICDE}}, pages 101--110, 2000.

\bibitem{ast}
M.~Zaharioudakis, R.~Cochrane, G.~Lapis, H.~Pirahesh, and M.~Urata.
\newblock Answering complex {SQL} queries using automatic summary tables.
\newblock {\em {VLDB}}, pages 105--116, 2000.

\end{thebibliography}

\clearpage
\appendix

\section{Existing JavaScript SQL Engines}
\label{sec:existing}

We are, of course, not the first to explore data management
inside the browser. Arguably, any web application that contains
JavaScript code and performs data manipulation requires some
data management capability. For example, popular JavaScript
libraries such as JQuery and D3.js provide basic filter queries over
JavaScript objects that model a web page (i.e., the DOM). Any direct
manipulation of JavaScript objects will have performance similar to
the J1 condition discussed in Section~\ref{sec:micro}, and as the
results in Section~\ref{eval:tpc} show, without typed arrays, the
amount of data that can be loaded in the browser is quite limited.
Any moderately-complex JavaScript visualization (in D3.js for
example) is likely to be performing aggregations, grouping, and even
joins---albeit in an ad hoc, imperative fashion. Nevertheless, a
direct performance comparison between Afterburner and JavaScript
libraries such as JQuery and D3.js would not be particularly
meaningful, since those libraries do not support SQL. Imperative
data processing libraries in JavaScript occupy a completely
different point in the design space.

There are, however, two existing SQL-in-JavaScript solutions
worth discussing:

\smallskip \noindent {\bf Lovefield}\footnote{\url{https://github.com/google/lovefield}}
is a relational database for web apps written by Google.  It is
implemented in JavaScript and runs entirely inside the browser. The
system does not support complex aggregates such as:

\smallskip
\begin{small}
\begin{Verbatim}[commandchars=\\\{\}]
  SUM(l_extendedprice * (1 - l_discount))
\end{Verbatim}
\end{small}
\smallskip

\noindent and therefore it is not capable of running Q1 from
TPC-H in its original form. While this is a rather minor issue, and
support for more complex queries can certainly be improved in
Lovefield, we do not expect the engine to scale and to achieve
performance comparable to Afterburner for a fundamental reason:
Lovefield's storage and execution engine depend on JavaScript
objects. Based on the comparison between the J1 and J3 conditions in
Section~\ref{sec:micro} and the results from Section~\ref{eval:tpc},
we know that this is a significant impediment to performance and scalability.

Nevertheless, to compare Afterburner with Lovefield, we generated an
instance of the {\tt lineitem} table using the TPC-H data generator
and incrementally ingested data into the system (on the same client
machine used in Section~\ref{sec:micro}). We were not able to load
more than the first 5000 records before the browser tab crashed. On
this amount of data, we attempted to run TPC-H Q1. Due to the
issues raised above, we had removed all unsupported expressions. For
concreteness, the following is the variant query expressed in
Lovefield's API:

\smallskip
\begin{small}
\begin{Verbatim}[commandchars=\\\{\}]
  tpch.select(LINEITEM.L_RETURNFLAG,
              LINEITEM.L_LINESTATUS,
              lf.fn.sum(LINEITEM.L_QUANTITY),
              lf.fn.sum(LINEITEM.L_EXTENDEDPRICE),
              lf.fn.avg(LINEITEM.L_QUANTITY),
              lf.fn.avg(LINEITEM.L_EXTENDEDPRICE),
              lf.fn.avg(LINEITEM.L_DISCOUNT),
              lf.fn.count(LINEITEM.L_ORDERKEY))
      .from(LINEITEM)
      .where(LINEITEM.L_SHIPDATE.lt(
             dbDatetolfDate('1998-09-02')))
      .groupBy(LINEITEM.L_RETURNFLAG,
               LINEITEM.L_LINESTATUS)
      .orderBy(LINEITEM.L_RETURNFLAG, lf.Order.ASC)
      .orderBy(LINEITEM.L_LINESTATUS, lf.Order.ASC)
      .exec()
\end{Verbatim}
\end{small}
\smallskip

\noindent Note that Lovefield uses a fluent approach to specifying
queries, just like Afterburner. The above query on 5000 records took
30ms. On the same query, Afterburner takes 11ms. It is important to note
that even with Afterburner's constant overheads per query (e.g., query
compilation, asm.js compilation, etc.), our system is still much
faster than Lovefield on this tiny dataset.

\smallskip \noindent {\bf
Sql.js}\footnote{\url{https://github.com/kripken/sql.js/}} is a
cross-compiled version of SQLite(v3) into asm.js.
It shares a few features with Afterburner:\ Sql.js
uses code generation but targets the SQLite virtual machine 
bytecode, employs asm.js (via cross compilation), uses in-memory
storage, and uses typed arrays for storage and query
execution (i.e., does not depend on JavaScript objects).

In our experiments, we were not able to load a TPC-H database of
scale factor 1GB into Sql.js (the browser hangs and then crashes).
We tried to load smaller slices of the database in 10\% increments.
We only managed to ingest the first 600k records from the 
\texttt{lineitem} table. For Q1, on this amount of data, Sql.js
took 3.55s to execute, compared to 22ms for Afterburner. We were unable
to test more complex queries such as multiple joins since they took
too long (more than one minute per query before we gave up).

Our evaluation shows that Sql.js is two orders of magnitude
slower than Afterburner for Q1. We attribute this performance gap to two
main drawbacks:\ First, SQLite's in-memory storage is not
optimized for analytical tasks (i.e., it is a row store as opposed to a column
store). Second, Afterburner's code generation is more efficient since it
generates asm.js directly, while Sql.js generates code that is again
interpreted in a cross-compiled virtual machine. While Sql.js aims
to be a full-fledged RDBMS, Afterburner has no such aspiration:\ we aim
to be a high-performance analytics RDBMS for a narrowly-targeted
application scenario and are able to optimize directly for this
usage.

\section{Performance Details}
\label{appendix:detail}

\begin{table*}[t]
\center
\begin{small}
\begin{tabular}[t]{|l||rr|rr|rr|rr|rr|rr|rr|}
\hline
&\multicolumn{2}{c|}{{\bf Afterburner}}&\multicolumn{2}{c|}{{\bf Afterburner}}&\multicolumn{2}{c|}{{\bf MonetDB}} &\multicolumn{2}{c|}{{\bf MonetDB}}&\multicolumn{6}{c|}{\bf LegoBase}  \\ \cline{10-15}
&\multicolumn{2}{c|}{{\bf (Full)}}&\multicolumn{2}{c|}{{\bf (Vanilla)}}&\multicolumn{2}{c|}{{\bf (Single-threaded)}}&\multicolumn{2}{c|}{{\bf (Multi-threaded)}}&
\multicolumn{2}{c|}{{\bf execute}}&\multicolumn{2}{c|}{{\bf codegen}}&\multicolumn{2}{c|}{{\bf compile}} \\ \hline\hline
Q01&115 &(8.0) &318 &(16.3)&1269&(10.6)&588 &(13.0)  &57  &(0.5) &159 &(35.4) &216& (0.6)\\ \hline
Q02&68  &(8.2) &137 &(14.8)&36  &(3.6)&20   &(0.0)   &27  &(0.6) &431 &(100.9)&351& (3.5)\\ \hline
Q03&108 &(4.4) &230 &(6.1)&242  &(80.7)&102 &(0.0)   &120 &(5.5) &328 &(89.1) &333& (1.4)\\ \hline
Q04&161 &(2.1) &361 &(30.3)&121 &(2.9)&60   &(0.0)   &163 &(13.5)&177 &(74.0)   &266& (4.5)\\ \hline
Q05&151 &(1.6) &385 &(13.1)&141 &(6.1)&72   &(0.0)   &73  &(2.9) &597 &(99.6) &501& (2.9)\\ \hline
Q06&36  &(0.4) &41  &(0.2)&431  &(25.7)&234 &(0.0)   &21  &(0.5) &67  &(24.9) &161& (11.4)\\ \hline
Q07&322 &(11.3)&595 &(18.8)&161 &(3.0)&90     &(0.0)   &\gc &\gc   &510 &(62.6) &525& (3.9)\\ \hline
Q08&101 &(2.9) &298 &(5.2)&113  &(3.0)&70     &(0.0)   &306 &(20.4)&1039&(66.3) &612& (3.9)\\ \hline
Q09&314 &(2.3) &635 &(20.6)&235 &(2.1)&122  &(3.9) &\gc &\gc   &710 &(94.6) &569& (26.4)\\ \hline
Q10&233 &(6.4) &528 &(33.5)&126 &(2.5)&70   &(0.0)   &336 &(2.0)   &382 &(21.7) &350& (2.6)\\ \hline
Q11&35  &(6.0)   &83  &(11.9)&42  &(1.8)&30   &(0.0)   &14  &(0.0)   &237 &(14.6) &297& (8.8)\\ \hline
Q12&82  &(1.4) &184 &(1.1)&119  &(2.7)&70   &(0.0)   &116 &(10.4)&192 &(11.0)   &267& (1.1)\\ \hline
Q13&830 &(2.4) &1065&(37.4)&243 &(4.9)&194  &(4.8) &56  &(0.0)   &131 &(11.1) &194& (1.4)\\ \hline
Q14&44  &(3.5) &70  &(1.4)&40   &(0.6)&20   &(0.0)   &25  &(1.0)   &123 &(9.6)  &218& (18.1)\\ \hline
Q15&50  &(1.0)   &101 &(0.9)&74   &(1.5)&98   &(3.9) &49  &(0.4) &106 &(6.8)  &213& (1.1)\\ \hline
Q16&149 &(3.2) &325 &(1.7)&316  &(7.0)&180    &(0.0)   &207 &(7.4) &433 &(18.1) &449& (4.9)\\ \hline
Q17&718 &(1.9) &1394&(57.2)&185 &(1.6)&140  &(10.7)&278 &(15.8)&156 &(7.2)  &271& (1.3)\\ \hline
Q18&446 &(69.6)&765 &(105.7)&171&(1.2)&162  &(0.0)   &414 &(0.8) &213 &(11.1) &261& (1.7)\\ \hline
Q19&1951&(62.0)  &2100&(139.3)&209&(1.6)&112  &(14.4)&175 &(4.6) &175 &(21.8) &288& (9.3)\\ \hline
Q20&62  &(0.6) &162 &(1.5)&86   &(2.7)&50   &(0.0)   &39  &(1.4) &436 &(20.6) &463& (50.2)\\ \hline
Q21&284 &(2.3) &564 &(39.0)&369   &(10.4)&228 &(9.6) &1112&(5.2) &371 &(14.9) &395& (40.4)\\ \hline
Q22&95  &(1.1) &156 &(2.3)&118  &(11.2)&92  &(3.9) &68  &(0.5) &241 &(28.7) &210& (0.9)\\ \hline
\hline
total*&\multicolumn{2}{r|}{5718}&\multicolumn{2}{r|}{9266}&\multicolumn{2}{r|}{4452} &\multicolumn{2}{r|}{2592}&\multicolumn{2}{r|}{3655}&\multicolumn{2}{r|}{5995}&\multicolumn{2}{r|}{6316}\\ \hline
total &\multicolumn{2}{r|}{6353}&\multicolumn{2}{r|}{10496}&\multicolumn{2}{r|}{4848}&\multicolumn{2}{r|}{2804}&\multicolumn{2}{r|}{\gc}&\multicolumn{2}{r|}{7215}&\multicolumn{2}{r|}{7410}\\ \hline
\end{tabular}
\vspace{0.2cm}
  \caption{Query latency (ms) for TPC-H queries, comparing Afterburner with MonetDB and LegoBase.
           Query latencies are averaged over five trials with 95\% confidence intervals.}
  \label{tab:allq}
\end{small}
\end{table*}

Table~\ref{tab:allq} provides detailed experimental results on all 22
TPC-H benchmark queries for our five experimental
conditions:\ Afterburner (Full), Afterburner (Vanilla), MonetDB
(Single-threaded), MonetDB (Multi-threaded), and LegoBase. These
results are summarized in Figure~\ref{fig:allq}, but the table
provides the exact values, along with 95\% confidence intervals. For
LegoBase, we show the codegen and compile phases as well as the actual
query execution latency. Note that Q7 and Q9 did not complete on our
client machine (not enough memory) with LegoBase.

\end{document}